\documentclass[a4paper,11pt]{article}
\pdfoutput=1 

\usepackage{jheppub} 
\usepackage{hyperref}
\usepackage{physics}
\usepackage{amsmath}
\usepackage{graphicx}
\usepackage{multirow}
\usepackage{graphics}
\usepackage{siunitx}
\usepackage{float}
\usepackage{subcaption}
\usepackage[titletoc]{appendix}
\usepackage{xspace}
\usepackage{tikz}
\floatstyle{plaintop}
\usepackage[export]{adjustbox}
\restylefloat{table}

\usepackage{listofitems} 
\usetikzlibrary{arrows.meta} 
\usepackage[outline]{contour} 
\contourlength{1.4pt}

\tikzset{>=latex} 
\usepackage{xcolor}
\colorlet{myred}{red!80!black}
\colorlet{myblue}{blue!80!black}
\colorlet{mygreen}{green!60!black}
\colorlet{myorange}{orange!70!red!60!black}
\colorlet{mydarkred}{red!30!black}
\colorlet{mydarkblue}{blue!40!black}
\colorlet{mydarkgreen}{green!30!black}
\tikzstyle{node}=[thick,circle,draw=myblue,minimum size=22,inner sep=0.5,outer sep=0.6]
\tikzstyle{node in}=[node,green!20!black,draw=mygreen!30!black,fill=mygreen!25]
\tikzstyle{node hidden}=[node,blue!20!black,draw=myblue!30!black,fill=myblue!20]
\tikzstyle{node convol}=[node,orange!20!black,draw=myorange!30!black,fill=myorange!20]
\tikzstyle{node out}=[node,red!20!black,draw=myred!30!black,fill=myred!20]
\tikzstyle{connect}=[thick,mydarkblue] 
\tikzstyle{connect arrow}=[-{Latex[length=4,width=3.5]},thick,mydarkblue,shorten <=0.5,shorten >=1]
\tikzset{ 
  node 1/.style={node in},
  node 2/.style={node hidden},
  node 3/.style={node out},
}
\def\nstyle{int(\lay<\Nnodlen?min(2,\lay):3)} 

\definecolor{codegreen}{rgb}{0,0.6,0}
\definecolor{codegray}{rgb}{0.5,0.5,0.5}
\definecolor{codepurple}{rgb}{0.58,0,0.82}
\definecolor{backcolour}{rgb}{0.95,0.95,0.92}
\definecolor{forestgreen}{RGB}{34,139,34}

\usepackage{enumitem}
\setlist{noitemsep}

\usepackage[T1]{fontenc} 

\usepackage[style=numeric-comp,sorting=none,backend=biber,maxbibnames=80]{biblatex}
\begin{document}


\renewcommand{\thefootnote}{\fnsymbol{footnote}}

\title{\boldmath Baler - Machine Learning Based Compression of Scientific Data \\}

\date{January 2023}

\author[1]{F. Bengtsson}
\author[2]{C. Doglioni}
\author[1]{P.A. Ekman}
\author[1]{A. Gall\'en}
\author[2]{P. Jawahar}
\author[1]{A. Orucevic-Alagic}
\author[2]{M. Camps Santasmasas}
\author[2]{N. Skidmore}
\author[2]{O. Woolland}

 \affiliation[1]{Lund University}
 \affiliation[2]{University of Manchester}

\abstract{Storing and sharing increasingly large datasets is a challenge across scientific research and industry. In this paper, we document the development and applications of Baler - a Machine Learning based data compression tool for use across scientific disciplines and industry. Here, we present Baler's performance for the compression of High Energy Physics (HEP) data, as well as its application to Computational Fluid Dynamics (CFD) toy data as a proof-of-principle. We also present suggestions for cross-disciplinary guidelines to enable feasibility studies for machine learning based compression for scientific data.}

\notoc
\maketitle
\thispagestyle{empty}
\pagenumbering{arabic} 
\section{Introduction}\label{introduction}
Many different fields of science share a common issue; storing ever-growing datasets. 
By the end of the next decade, the Large Hadron Collider (LHC) experiments will have over an order of magnitude more data to analyze than currently~\cite{atlas, cms, lhcb}; the Square Kilometre Array (SKA) experiment is expected to record 8.5EB of data over its 15-year lifespan~\cite{SKA} and fields such as Computational Fluid Dynamics (CFD) rely on TB-sized simulation samples that need to be stored and shared. 
Without significant R\&D, the datasets expected to be collected by big-data science experiments are projected to exceed the available storage resources (see e.g. Fig.~2 of Ref.~\cite{atlas} for the case of the ATLAS experiment at the LHC). 
This cross-disciplinary issue is not limited to scientific research and extends to industrial operations~\cite{7996801}.

\subsection{Lossy data compression in high energy physics}

A common mitigation strategy to this problem involves compressing data using lossless algorithms, see e.g. Refs.~\cite{Shadura_2020, Patauner:1433839, Rawal:2020}. 
Once the storage limit is reached, one is forced to discard parts of the dataset, or only save certain features of the data. 
Generally, this can be done without impacting the overall scientific program of the experiments, for example by using a data selection system called \textit{trigger} that only stores data satisfying certain pre-determined characteristics that ensure the dataset will be aligned with the experiment's main scientific goals. 
However, saving only a subset of data is not ideal for processes where additional statistical power is necessary, e.g. for rare signals buried in high-rate backgrounds. 
In these cases, one can foresee using \textit{lossy} compression algorithms that reduce the data size ideally beyond what lossless compression algorithms can do~\cite{sayood2017introduction}, using approximation and partial data discarding, at the expense of data fidelity. 
One limitation of lossy compression is that to obtain high compression ratios with low data loss, the compression algorithm must be tailored to the input data; for instance, MP3~\cite{brandenburg1999mp3} is an example of a lossy compression algorithm that uses techniques specifically suited for waves and frequencies. 
Thereby, a general solution to this cross-disciplinary problem is hard to obtain by traditional methods. 
As a solution to this problem, we present Baler, a lossy data compression tool based on the machine learning autoencoder architecture, which tailors the compression to the user's dataset.
It is also important to note that for such a tool to be usable in a scientific experiment, the loss in data quality must be controlled and it must also be deemed to be tolerable/negligible with respect to other sources of experimental jitter. 

\subsection{Autoencoders for lossy data compression}
\begin{figure}[H]
\centering
\begin{tikzpicture}[x=1.7cm,y=0.9cm]
      \large
      \message{^^JNeural network without arrows}
      \readlist\Nnod{4,5,4,2,4,5,4} 

      \node[above,align=center,myorange!60!black] at (3,2.4) {encoder};
      \node[above,align=center,myblue!60!black] at (5,2.4) {decoder};
      \node[above,align=center,black] at (4,1.9) {latent space};
      \draw[myorange!40,fill=myorange,fill opacity=0.02,rounded corners=2]
        (1.6,-2.7) --++ (0,5.4) --++ (2.8,-1.2) --++ (0,-3) -- cycle;
      \draw[myblue!40,fill=myblue,fill opacity=0.02,rounded corners=2]
        (6.4,-2.7) --++ (0,5.4) --++ (-2.8,-1.2) --++ (0,-3) -- cycle;
      
      \message{^^J  Layer}
      \foreachitem \N \in \Nnod{ 
        \def\lay{\Ncnt} 
        \pgfmathsetmacro\prev{int(\Ncnt-1)} 
        \message{\lay,}
        \foreach \i [evaluate={\y=\N/2-\i+0.5; \x=\lay; \n=\nstyle;}] in {1,...,\N}{ 
          
          \node[node \n,outer sep=0.6] (N\lay-\i) at (\x,\y) {};
          
          \ifnum\lay>1 
            \foreach \j in {1,...,\Nnod[\prev]}{ 
              \draw[connect,white,line width=1.2] (N\prev-\j) -- (N\lay-\i);
              \draw[connect] (N\prev-\j) -- (N\lay-\i);
            }
          \fi 
          
        }
      }
      
      \node[above=2,align=center,mygreen!60!black] at (N1-1.90) {input};
      \node[above=2,align=center,myred!60!black] at (N\Nnodlen-1.90) {output};
    \end{tikzpicture}
    \caption{Illustration of an autoencoder consisting of an input and output layer. In between the input and output, there are hidden layers and a latent space, where the dimensionality of the latent space is less, equal, or greater to the input and output layers. Modified from Ref.~\parencite{NeuralNetworks}.}
    \label{fig:autoencoder}
\end{figure}
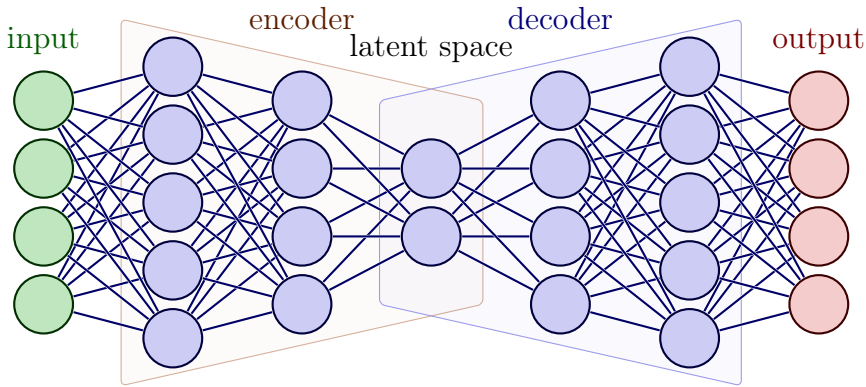

Autoencoders (AEs)~\cite{Kramer1991NonlinearPC} are a class of unsupervised deep neural networks characterized by an encoder, a central latent space, a decoder, and a target space of the same dimensionality as the input space, as illustrated in Figure~\ref{fig:autoencoder}. The encoder, is a neural network that maps each input, $x$, to an abstract latent point $z$, generally of lower dimensionality than the input. The decoder then extrapolates the latent space back to the same dimensions as the input to give the reconstructed output, $\hat{x}$. AEs can therefore be trained to reconstruct the various features of the input data, while their bottleneck structure prevents them from simply learning the identity map. The dimensionality of the latent space is of particular importance, as it determines the amount of compression achieved, with the latent points being the compressed data and the decoder acting as the decompression algorithm.
AE based compression of scientific data has shown promising results for multiple fields of study such as meteorology, cosmology, computational fluid dynamics, crystallography etc.~\cite{9380370,9555941,wang2022locality,la2022hyperspectral,9680154,lee2022error,sriram2022deepcomp}. 
The use of AEs for data compression in High Energy Physics (HEP) has been also shown to be promising in previous studies~\cite{9004751,9012882,Collins:2022qpr,LHCbCompression}. 
A number of these studies focus on the compression of objects directly as the data is taken (\textit{online} compression), which would technically require to train a model on a dataset and using it to compress a different dataset with the same input characteristics.
\textit{Offline} compression on the other hand corresponds to the case where the model is trained to compress a dataset and is used to compress that dataset only. 
In this work, we deal with offline compression as a stepping stone toward online compression and leave the latter for future studies.

\section{Baler methodology}
\label{sec:method}

AE based compression workflows generally consist of data pre-processing, model architecture selection, model training, compression and decompression using a trained model, and performance evaluation via selected metrics. Baler is an intuitively packaged, modular tool that allows for easy modifications in any component of the workflow. Baler is available in its open-source software repository~\cite{baler}. In this section, we discuss the setup for HEP data compression as a working example.

\subsection{ HEP model design}\label{Design}

The benchmark HEP AE is built with $3$ fully connected neural network layers in both the encoder and decoder. The encoder layers have 200, 100, and 50 nodes respectively, while the decoder has a symmetrically inverted layer structure. We train our models by minimizing the loss function:
\begin{equation}
    L_{total} = (1-\beta) L_{reco} + \beta L_{sparse}
    \label{eqn:loss}
\end{equation}
where, $\beta$ is a free hyperparameter that controls the contribution of each term to the net loss, $L_{reco}$ is a suitable reconstruction error metric and $L_{sparse}$ is a $L_1$-type regularization term to enforce sparsity in the AE weights. In this work we choose $L_{reco}$ to be the mean squared error (MSE) summed over each mini-batch, defined as,
\begin{equation*}
    \text{MSE} = \frac{1}{n} \sum^m_{j=1}\sum^n_{i=1} (X_i - \hat{X}_i)_j^2 \label{eqn:mse}
\end{equation*}
Here, $m$ is the batch-size and $n$ is the number of variables in each data entry. $X$ is the vector of batched model inputs, and $\hat{X}$ is the vector corresponding model reconstruction. $L_{sparse}$ introduces sparsity to the model and reduces its storage size, thereby reducing the overheads as discussed in Sec.~\ref{Training setup and evaluation metrics}. $L_{sparse}$ is defined as,
\begin{equation}
    \text{L}_{sparse} = \sum_i \abs{w_i}
    \label{eqn:L1}
\end{equation}
and has previously been shown to perform better with regards to HEP data compression~\cite{dialektakis_george_2021_5482611}.

\subsection{HEP training setup and evaluation metrics}\label{Training setup and evaluation metrics}
The models are built using the \texttt{PyTorch}~\cite{Pytorch} framework and optimized using the Adam minimizer~\cite{AdamOptimizer}. A learning rate of $\SI{e-3}{}$ is used in combination with a learning rate scheduler, namely the \texttt{ReduceLROnPlateau} method from \texttt{PyTorch}. The scheduler uses a patience of $50$ epochs, a reduction factor of $0.5$, and a minimum learning rate of $\SI{e-6}{}$. We train for $1000$ epochs with a batch size of $512$, and an early stopping strategy with a patience of $100$ epochs. $L_{total}$ converges to an order of $\SI{e-5}{}$.

We choose the residual and response as our performance evaluation metrics, defined as,
\begin{align}
    \text{residual} &= \hat{X} - X
    \label{eq:residual}
    \\
    \text{response} &= \frac{\text{residual}}{X},
    \label{eq:response}
\end{align}
with $X$ being the original data and $\hat{X}$ being the data reconstructed from the compressed file. Another important metric for compression is the compression ratio, defined as,
\begin{equation}
    R = \frac{\mathrm{size}(\text{input file})}{\mathrm{size}(\text{compressed file})}.
\end{equation}
However, unlike some traditional compression algorithms, AE compression requires auxiliary files to be saved. Auxiliary files mainly include decoder weights and biases along with the corresponding \texttt{PyTorch} metadata required to load the model when decompressing. This information is saved using the save functionality provided by \texttt{PyTorch}. Auxiliary files can also include auxiliary data such as normalization features and data headers. Taking this into account, the actual compression ratio is 
\begin{equation}
    R_{\text{actual}} = \frac{\mathrm{size}(\text{input file})}{\mathrm{size}(\text{decompressed file} + \text{auxiliary file(s)})}
\end{equation}

To avoid the results being skewed due to a single specific seed being used in the training, the training was done using 10 different seeds, and the performance evaluation was done on the 5 best performing seeds. For the offline compression case studied here, choosing the best seed is considered as a type of hyperparameter optimization. The impact of different seeds will be studied in the future.

\subsection{HEP implementation}

For the results presented in this work, Baler release v1.0.0~\cite{zenodo_baler} was used. The computations used to obtain the results presented in this work were performed on the AURORA cluster hosted by LUNARC~\cite{AuroraLunarc} at Lund University. The nodes on the cluster are running CentOS 7.2 x86\_64 and consist of 2 Intel Xeon E5-2650 v3 ($\SI{2.3}{GHz}$, 10-core) with 64 GB of Memory (3.2 GB/core). By using the CUDA implementation in \texttt{PyTorch} the training of the AE was done using NVIDIA TESLA K80 GPUs. The prototyping and developing of the GPU implementation of Baler were performed with the help of the Blackett Computing Facility~\cite{BlackettUoM} at the University of Manchester.

\section{Baler input data}\label{Data Preparation}

Baler supports \texttt{NumPy}~\cite{numpy} arrays as input and output. 
This format was chosen because \texttt{NumPy} arrays are an easy-to-handle data format that is already widely used across various scientific disciplines. Also, since the \texttt{PyTorch}~\cite{Pytorch} library at the core of Baler uses tensors and conversion from the user's original file format is necessary and simple with \texttt{NumPy} arrays. 
In this initial study, we will focus on HEP data, and touch on preliminary studies using data from CFD. 

\subsection{HEP input data}

Processes involving the strong force dominate proton-proton collision interactions at the LHC. 
Therefore, one of the most commonly occurring observable objects at experiments like ATLAS and CMS are the collimated showers of particles resulting from these strong processes, reconstructed into \textit{jets}~\cite{Salam:2020lcn}. 

To showcase Baler's performance on HEP data, we use a subset of the jet data recorded by the CMS experiment at the LHC in 2012, released as open data under the Creative Commons CC0 waiver~\cite{dataset}. 
In this dataset, each entry is a proton-proton collision \textit{event}, and each event can contain multiple jets. 
Each collision event is independent from other events and there is no time-dependency in this data. 
In the data, jets are represented as 4-vectors ($p_T$, $\eta$, $\phi$, $m$). Where $p_T$ is the momentum of the jet perpendicular to the direction of the colliding proton beams, $\eta$ is a quantity related to the angle between the particle momentum and the beam, $\phi$ is the azimuthal angle measured around the beam axis, and $m$ is the mass of the jet. Collectively, these variables are called the jet's four-momentum which are the most relevant variables for LHC measurements and analyses involving jets. 
Each jet has several other associated variables. 
For example "jet area" is a measure of the footprint size of the jet.
The full list of variables and further information about the content of the dataset we use for testing Baler can be found in Ref.~\cite{CMSPhys2015}. 
At this stage, it is not clear whether it would be recommended to use a lossy compression algorithm on the four-momenta, but we include them in the bench-marking of the algorithm for this initial study.   

\subsection{HEP data pre-processing}

For a simpler use of the HEP data in Baler, the data is pre-processed. First, the data is flattened as the original hierarchical data structures of the input data are not supported by current machine learning frameworks such as \texttt{PyTorch}. 
This means that each jet in the dataset is considered independently from the others and correlations within events are lost\footnote{For early studies of the impact of considering individual jets versus jets from the same event, see Ref.~\cite{9075881}}. 
Secondly, features of the data which are non-numerical, are dropped. 
Both these pre-processing steps are limitations in the applicability of this compression method that can be overcome at a later stage. 

This pre-processing step removes nine variables only containing zeroes, and further truncates $15\%$ of the data, yielding a final dataset with a size of $\SI{116.9}{MB}$ consisting of $24$ variables and $608,978$ entries. A list of the $24$ variables can be found in Appendix \ref{lab:Results_table_appendix}.

\section{Baler performance on HEP data}\label{Results}

As described in Section \ref{Training setup and evaluation metrics}, we perform multiple training runs on the same dataset with different random seeds to account for statistical variations introduced in the model by seeds. We visualize the performance of Baler by looking a the 4-momentum variable distributions and their response distribution for one seed, as well as the mean and root mean squared (RMS) values of the response and residual distributions averaged across the five different seeds.

\begin{figure}[H]
    \centering
    \includegraphics[page=1,width=1\linewidth]{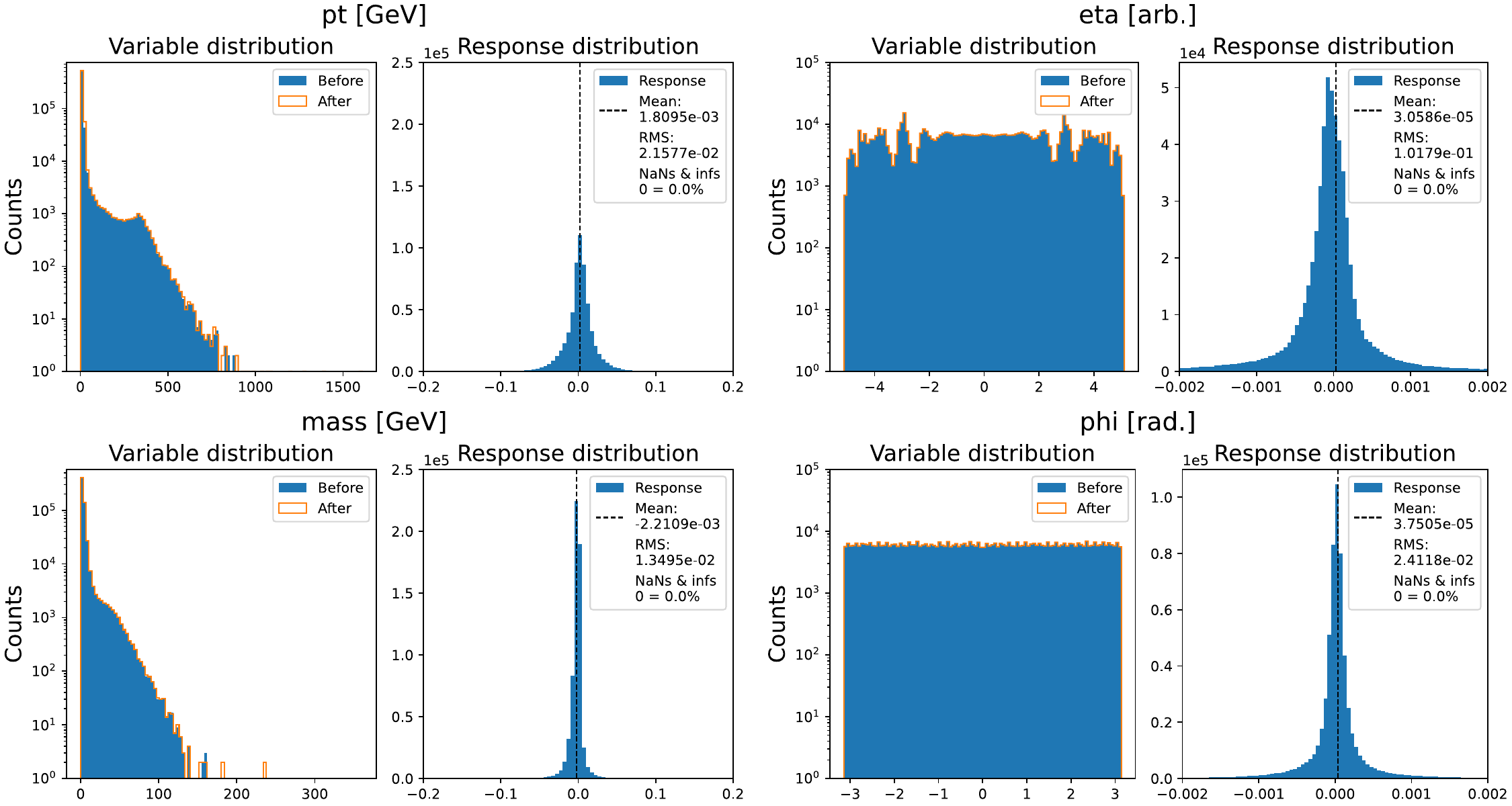}
    \caption{Distributions of the variables making up the 4-momentum of the jets. Alongside each distribution is a histogram of the response for that variable after compression with $R = 1.7$.}
    \label{fig:4momentum_performance}
\end{figure}

\begin{figure}[H]
    \centering
    \includegraphics[page=1,width=1\linewidth]{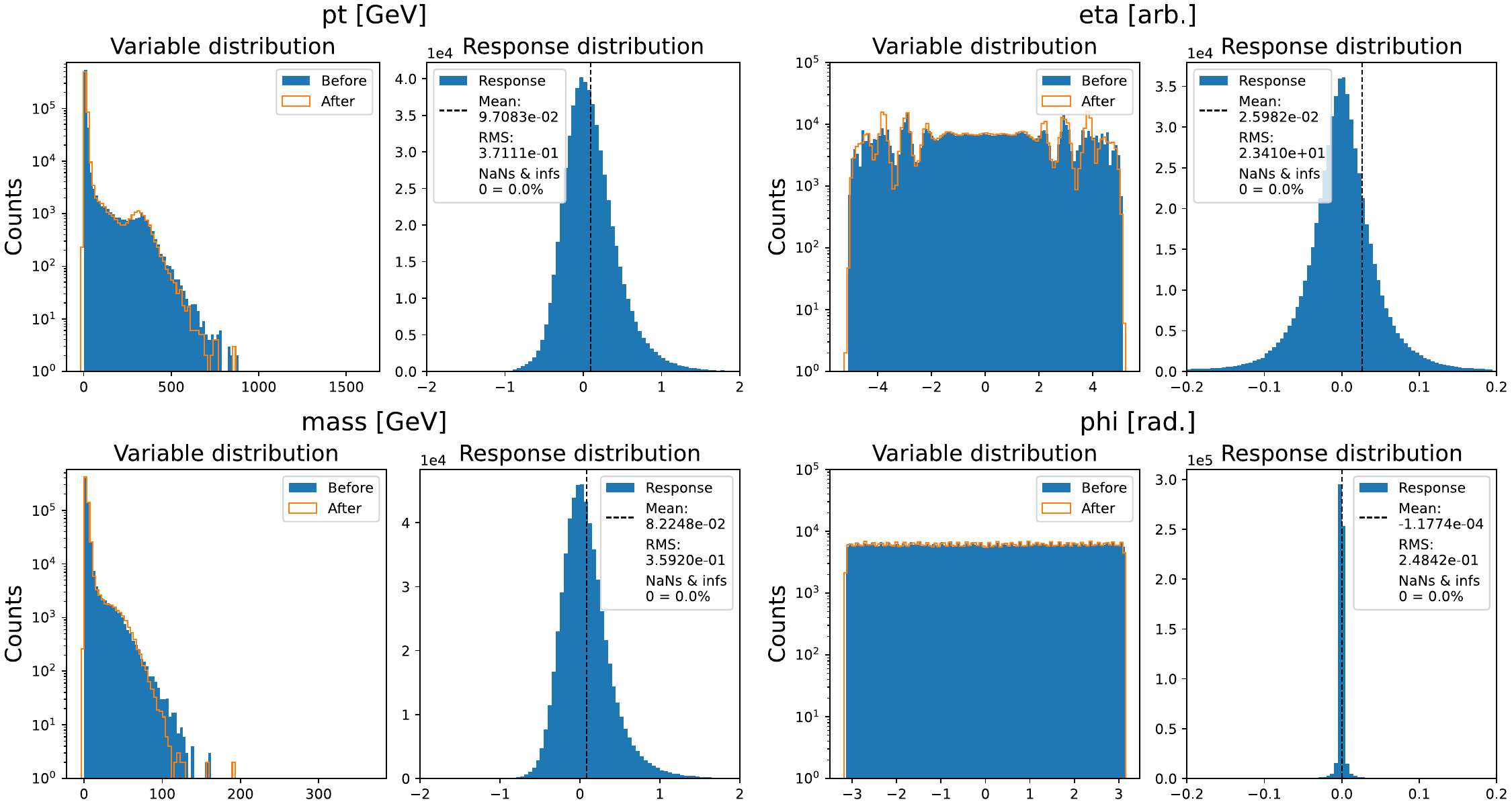}
    \caption{Distributions of the variables making up the 4-momentum of the jets. Alongside each distribution is a histogram of the response for that variable after compression with $R=6$.}
    \label{fig:4momentum_performance_6}
\end{figure}

Figure~\ref{fig:4momentum_performance} shows, for one seed, the distribution of the 4-momentum variables after compression and decompression with $R = 1.7$. Figure~\ref{fig:4momentum_performance_6} shows the corresponding plot with $R=6$. Alongside each distribution is a histogram of the response, defined in Eq. \ref{eq:response}, for that variable. These figures show that Baler achieves good reconstruction of the overall distributions whilst retaining a tight distribution of jet-by-jet variation for these variables at $R = 1.7$, whilst $R=6$ shows wider distributions and worse overall reconstruction. For example, Figure~\ref{fig:4momentum_performance} shows that at $R = 1.7$ the average $p_T$ response is 0.0018. The remaining 20 variables are presented in Appendix~\ref{app:distributions}.

Regarding the 4-momentum variables only, Table~\ref{tab:4vec_values} shows the mean and RMS of the residual and response for 5 different seeded runs at $R = 1.7$ and $R=6$, with an added statistical error of two standard deviations. Appendix~\ref{lab:Results_table_appendix} shows the same for the remaining 20 variables.


The difference between $R$ and $R_{\text{actual}}$ for HEP data is negligible. The model size on disk reaches approximately $\SI{500}{KB}$ and, including normalization features plus headers, the total auxiliary file size doesn't exceed $\SI{550}{KB}$. This means that for our case, where our input file size is $\SI{116.9}{MB}$, we obtain: $R = 1.7 \to R_{\text{actual}} \approx 1.59$, and $R = 6 \to R_{\text{actual}} \approx 5.8$. As the auxiliary file sizes for HEP do not increase with the number of entries they become negligible.

\begin{table}[H]
    \centering
    \caption{Residual and response distribution means and RMS values for the 4-momentum variables. These values are presented at two different compression ratios, and all values have been averaged over $5$ runs of different random seeds.}
    \setlength{\tabcolsep}{6pt} 
    \renewcommand{\arraystretch}{1} 
    \resizebox{1\linewidth}{!}{\begin{tabular}{|c|c|c|c|c|c|}
        \hline
        \multirow{2}{*}{Variable} & \multirow{2}{*}{Metric} & \multicolumn{2}{c|}{$R = 1.7$} & \multicolumn{2}{c|}{$R = 6$}\\
        \cline{3-6}
           & & Mean & RMS & Mean & RMS\\
        \hline
          \multirow{2}{*}{$p_T$} &Residual  &  $\SI{-1.44e-02}{} \pm \SI{1.04e-01}{}$ & $\SI{2.12e-01}{} \pm \SI{5.29e-02}{}$  & $\SI{-5.60e-02}{} \pm \SI{1.52e-01}{}$ & $\SI{1.17e+01}{} \pm \SI{3.13e+00}{}$ \\ \cline{2-6}
          &Response & $\SI{-1.07e-03}{}\ \pm \SI{1.34e-02}{}$ & $\SI{2.09e-02}{} \pm \SI{3.56e-03}{}$ & $\SI{9.08e-02}{}\ \pm \SI{2.37e-02}{}$ & $\SI{3.67e-01}{} \pm \SI{4.17e-02}{}$ \\ \hline   

          \multirow{2}{*}{$\eta$} &Residual  &  $\SI{-1.12e-03}{} \pm \SI{2.67e-03}{}$ & $\SI{2.09e-03}{} \pm \SI{1.45e-03}{}$ & $\SI{-2.14e-03}{} \pm \SI{6.21e-03}{}$ & $\SI{1.47e-01}{} \pm \SI{3.67e-02}{}$ \\ \cline{2-6} 
          &Response & $\SI{3.75e-04}{}\ \pm \SI{6.11e-04}{}$ & $\SI{8.12e-01}{} \pm \SI{1.17e+00}{}$ & $\SI{-5.42e-02}{}\ \pm \SI{3.34e-01}{}$ & $\SI{8.28e+01}{} \pm \SI{1.32e+02}{}$  \\ \hline
          
          \multirow{2}{*}{$\phi$} &Residual  &  $\SI{2.45e-04}{} \pm \SI{1.80e-03}{}$ & $\SI{9.91e-04}{} \pm \SI{1.12e-03}{}$ & $\SI{2.52e-04}{} \pm \SI{1.46e-03}{}$ & $\SI{9.92e-03}{} \pm \SI{2.12e-02}{}$\\ \cline{2-6} 
          &Response & $\SI{3.44e-04}{}\ \pm \SI{8.64e-04}{}$ & $\SI{1.93e-01}{} \pm \SI{4.32e-01}{}$ & $\SI{1.14e-04}{}\ \pm \SI{1.52e-03}{}$ & $\SI{6.63e-01}{} \pm \SI{8.32e-01}{}$ \\ \hline

          \multirow{2}{*}{mass} & Residual  &  $\SI{-8.05e-03}{} \pm \SI{2.51e-02}{}$ & $\SI{3.98e-02}{} \pm \SI{1.42e-02}{}$ & $\SI{1.22e-02}{} \pm \SI{2.55e-02}{}$ & $\SI{1.86e+00}{} \pm \SI{1.94e-01}{}$ \\ \cline{2-6} 
          &Response & $\SI{2.39e-01}{}\ \pm \SI{7.87e+00}{}$ & $\SI{4.38e+03}{} \pm \SI{4.47e+03}{}$ & $\SI{-1.34e+01}{}\ \pm \SI{5.05e+01}{}$ & $\SI{5.95e+04}{} \pm \SI{3.42e+04}{}$ \\ \hline
          \end{tabular}}
          \label{tab:4vec_values}
    \end{table}

\section{Baler application in other scientific fields}\label{other_fields}
Since a major goal of Baler is to investigate the feasibility of AE compression in different fields of science, Baler was also tested on simulated toy data from CFD. The simulated dataset used for this test was the x-component velocity of air flowing over a cube mounted to a wall. For simplicity, we only considered one slice in 3D space, making the compression of the 2D data simple using a convolutional-AE model where the encoder and decoder are Convolutional Neural Networks~\cite{lecun1995convolutional}.

\begin{figure}[H]
  \begin{subfigure}{0.31\textwidth}
    \includegraphics[width=\linewidth]{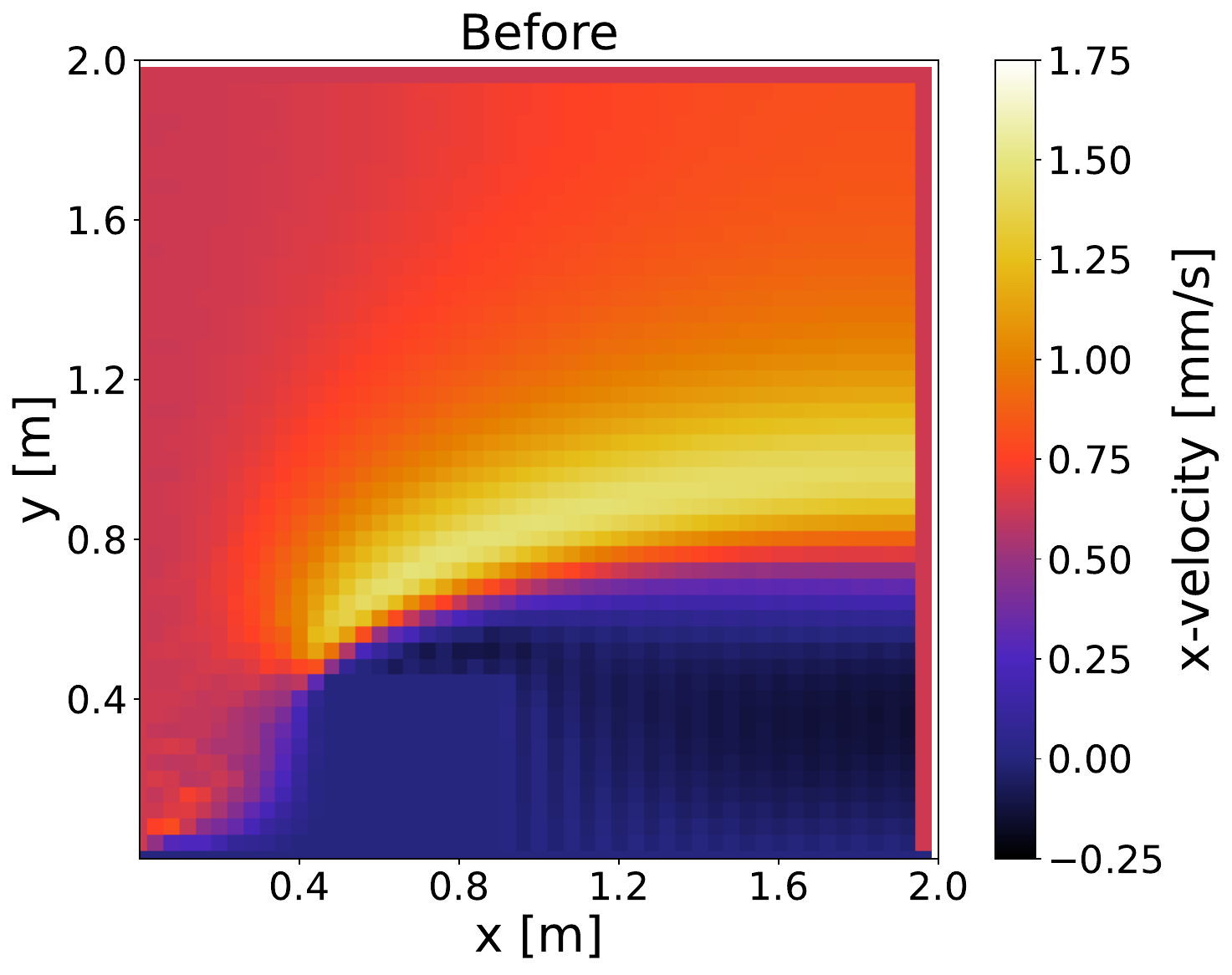}
    \caption{} \label{fig:CFD_before}
  \end{subfigure}%
  \hspace*{\fill}   
  \begin{subfigure}{0.31\textwidth}
    \includegraphics[width=\linewidth]{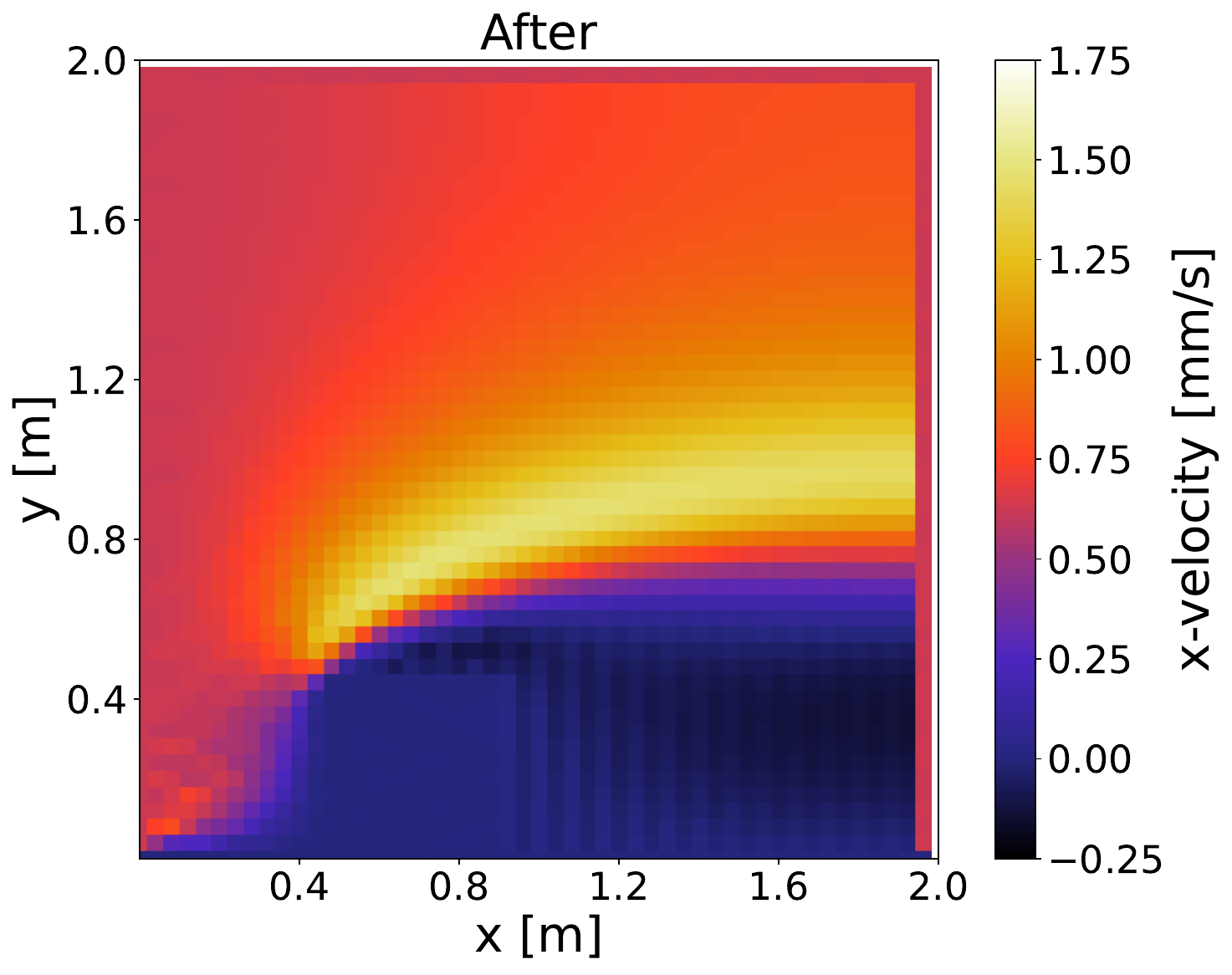}
    \caption{} \label{fig:CFD_after}
  \end{subfigure}%
  \hspace*{\fill}   
  \begin{subfigure}{0.295\textwidth}
    \includegraphics[width=\linewidth]{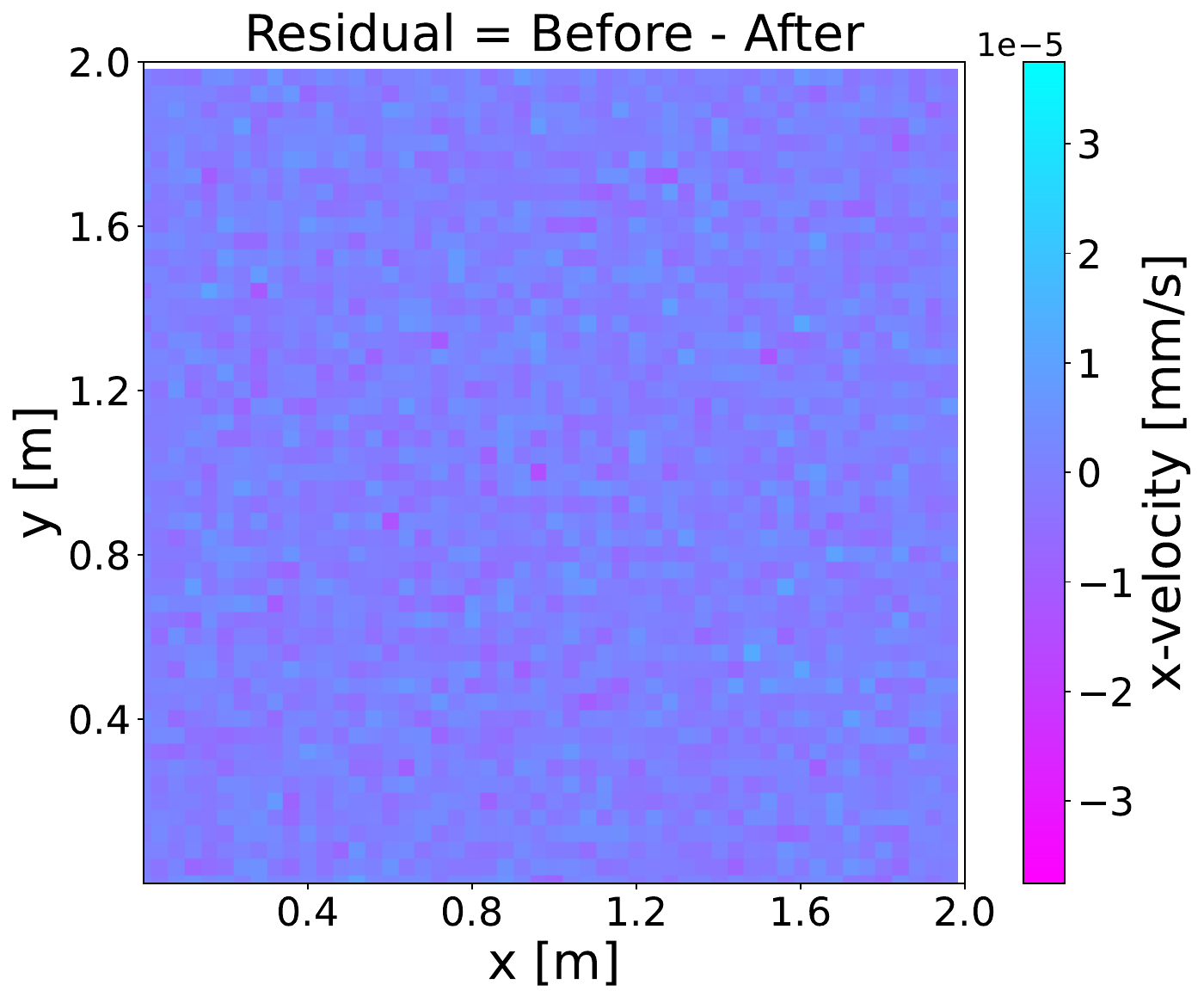}
    \caption{} \label{fig:CFD_diff}
  \end{subfigure}

\caption{A Computational Fluid Dynamics toy simulation showing x-component of air velocity before compression with $R = 88$ (a), after decompression (b), and the difference between the two (c).} \label{fig:CFD_full}
\end{figure}

Figure~\ref{fig:CFD_before} and~\ref{fig:CFD_after} show the 2D data before and after compression and decompression, with $R = 88$. Figure~\ref{fig:CFD_diff} shows the difference between the two which is on a scale four orders of magnitude smaller. As these results show Baler's wider applications to multiple scientific disciplines.

\section{Comparison to gzip}
Previous studies report a typical \texttt{gzip}~\cite{gzip} compression ratio of 1.5 - 2 for scientific data~\cite{IEEE_paper,vsgzip}. It has also been shown that an AE based compression can reach compression ratios of 100-500 with a point-wise relative error bound to 0.01~\cite{IEEE_paper}.

As shown in Figure~\ref{fig:4momentum_performance} and Figure~\ref{fig:4momentum_performance_6}, Baler can compress and decompress the HEP data with $R = 1.7$ more accurately than $R=6.0$. However, for this specific dataset, \texttt{gzip} can perform a lossless compression of $R=4.0$. Furthermore, applying \texttt{gzip} to the Baler output, which is already compressed to $R = 1.7$, does not significantly reduce the size of the Baler output. This is because as Baler decreases the dimensionality of the dataset, very few repeating values remain once compressed, and therefore methods like \texttt{gzip} are ineffective at compressing the Baler output further. This behavior is true even for larger file sizes: 250, 500, 750, 1000, 1250, and 1500 MB, as shown in Appendix~\ref{app:baler_vs_zip}.

On the contrary, for the CFD data shown in Figure~\ref{fig:CFD_full}, \texttt{gzip} achieves $R=2.2$ whilst Baler achieves $R=88$. The dilemma here is that the decoder Baler produces as part of the auxiliary files is 0.6 GB, making $R_{\text{actual}} < 1$. This is however not a major concern, since other studies have shown good performance with respect to the overhead~\cite{IEEE_paper}, and we expect to achieve similar lightweight models in the future. It is also important to note that the CFD dataset used in this work is a small toy since this is a proof-of-concept study. With AE compression, a model of fixed size can compress both small and relatively larger datasets, with the model overhead being large in the first case and much smaller in the second. We see this attenuating model overhead trend with HEP data, and we expect to see the same for CFD data, pending further studies with larger datasets.

\section{Conclusions and Outlook}\label{conclusions}

In this work, we motivate the need for effective data compression strategies as a solution to the growing storage issues related to large data volumes across many disciplines of scientific research. We present Baler as a modular solution to leverage machine learning based lossy data compression. We identify and define two major use cases of the tool namely, online and offline data compression. We evaluate performance for offline compression of HEP data using autoencoders and provide a guide to performing similar feasibility studies for other disciplines of research using the tool we have developed. We also study the compression of CFD data as a proof-of-concept to demonstrate Baler's flexibility.



Near future extensions to this work include assessing performance variations related to dataset sizes and support for error-bound compression where a specified error metric is used as a limit on the maximum possible compression ratio. Though we provide guidelines for using Baler to perform feasibility studies for a given dataset, there is currently no method to quickly project the likelihood of a dataset being suitable for compression with Baler. A potential way to implement this is to calculate a coefficient of variation for a given dataset as described in \cite{IEEE_paper} and use this as a likelihood metric. This implementation along with studies on other potential solutions for this problem are marked as features for future releases of Baler.

To deal with variations across dataset sizes we plan to perform follow-up studies by exploring different HEP datasets and input representations that may involve the use of low-level detector data.
Another related limitation is compressing files larger than RAM, where we plan to test industry standards such as optimal caching of objects in memory. These solutions are viable for offline compression since there are no associated latency or resource constraints in this case.

However, online compression is a major area of study we intend to carry out given its high potential to be used in large HEP experiments that generate data at very high rates. To tackle the problem of online compression we would need better generalization capabilities within the machine learning models and a potential way to achieve this with unsupervised learning is to use probabilistic generative models such as variational autoencoders and normalizing flows.

\raggedright
\clearpage

\acknowledgments{ This work was supported by the Turing-Manchester Feasibility Project Funding. PJ is part of SMARTHEP which received funding from the European Union’s Horizon 2020 research and innovation program under Grant Agreement n. 956086.}

\printbibliography[title={References},heading=bibintoc]

\newpage
\appendix
\section{Additional Plots}\label{appendix}
\subsection{Variable distribution plots }
\label{app:distributions}
Here, the distribution of the 24 variables compressed is displayed against their decompressed counterpart at different compression ratios. Note that the definition of the response leads to inf and NaN values which are not represented in these response distributions. The extent of this is presented in the legend of the following figures.

\begin{figure}[H]
    \centering
    \includegraphics[page=2,width=0.85\linewidth]{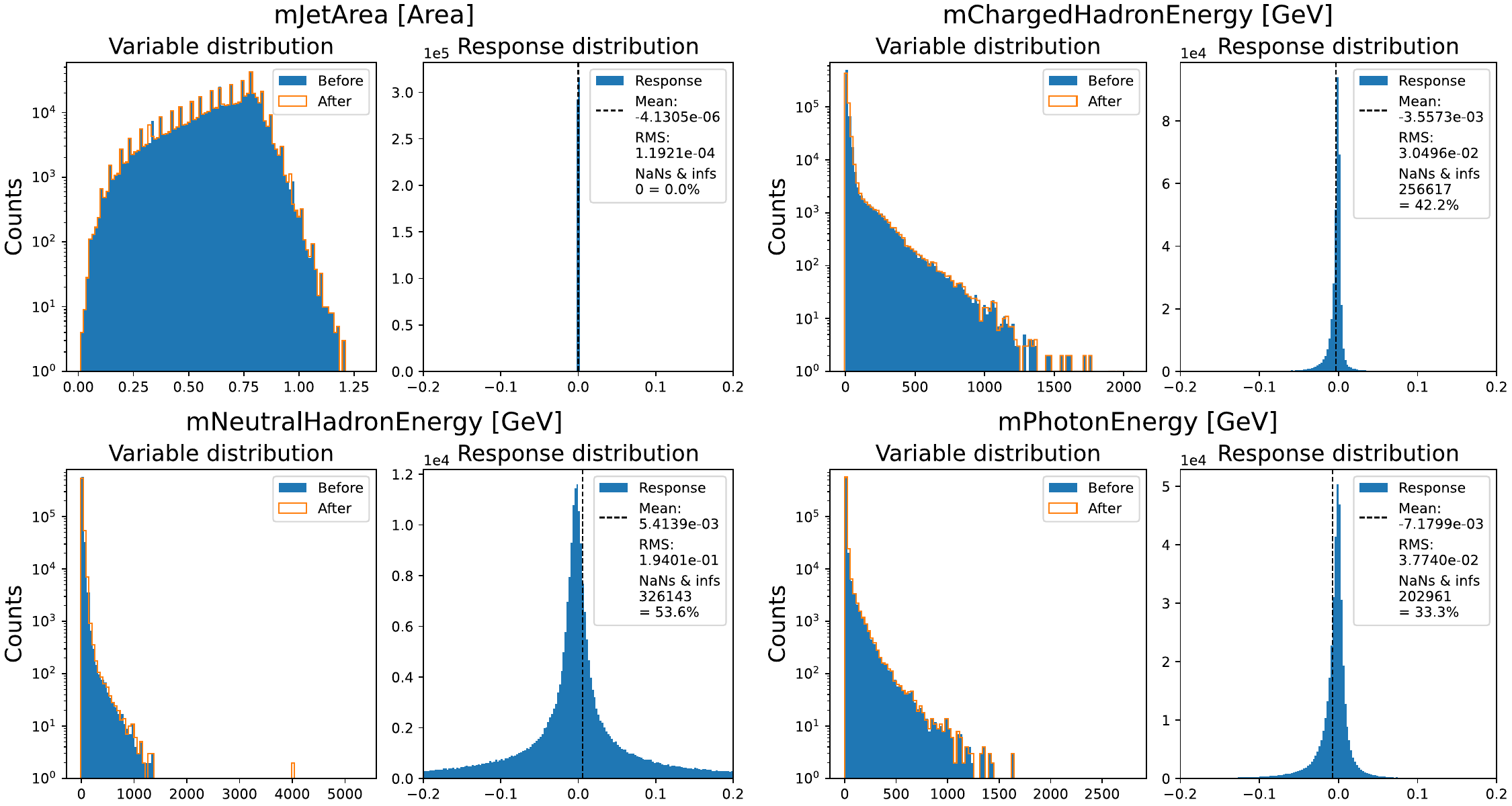}
    \includegraphics[page=3,width=0.85\linewidth]{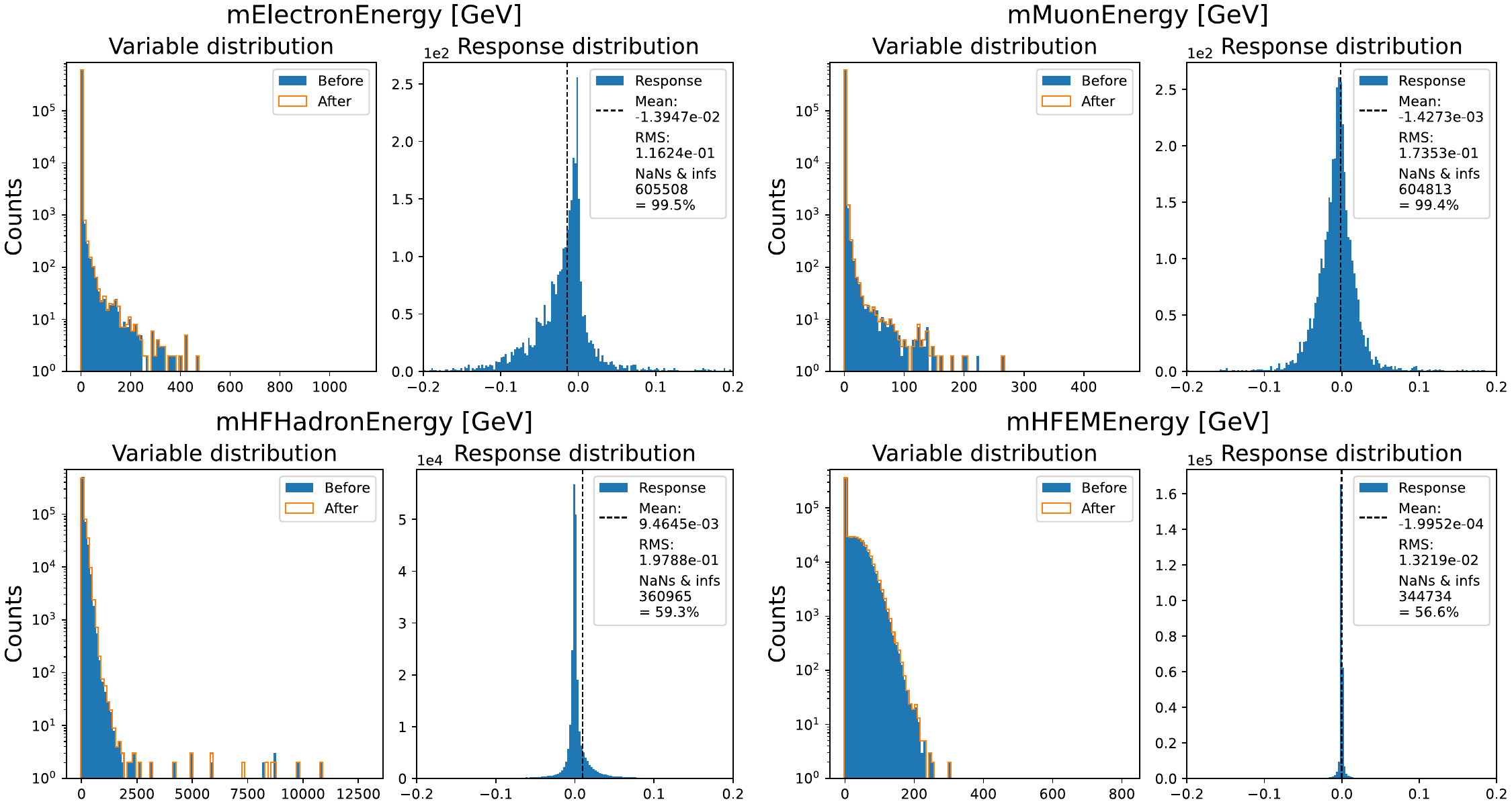}

    \caption{Original distribution and reconstructed distributions of eight jet variables after compression with $R = 1.7$. Alongside each distribution is a histogram of the response for that variable.}
    \label{fig:beforeafter1.6x}
\end{figure}
\begin{figure}[H]
    \centering
    \includegraphics[page=4,width=0.85\linewidth]{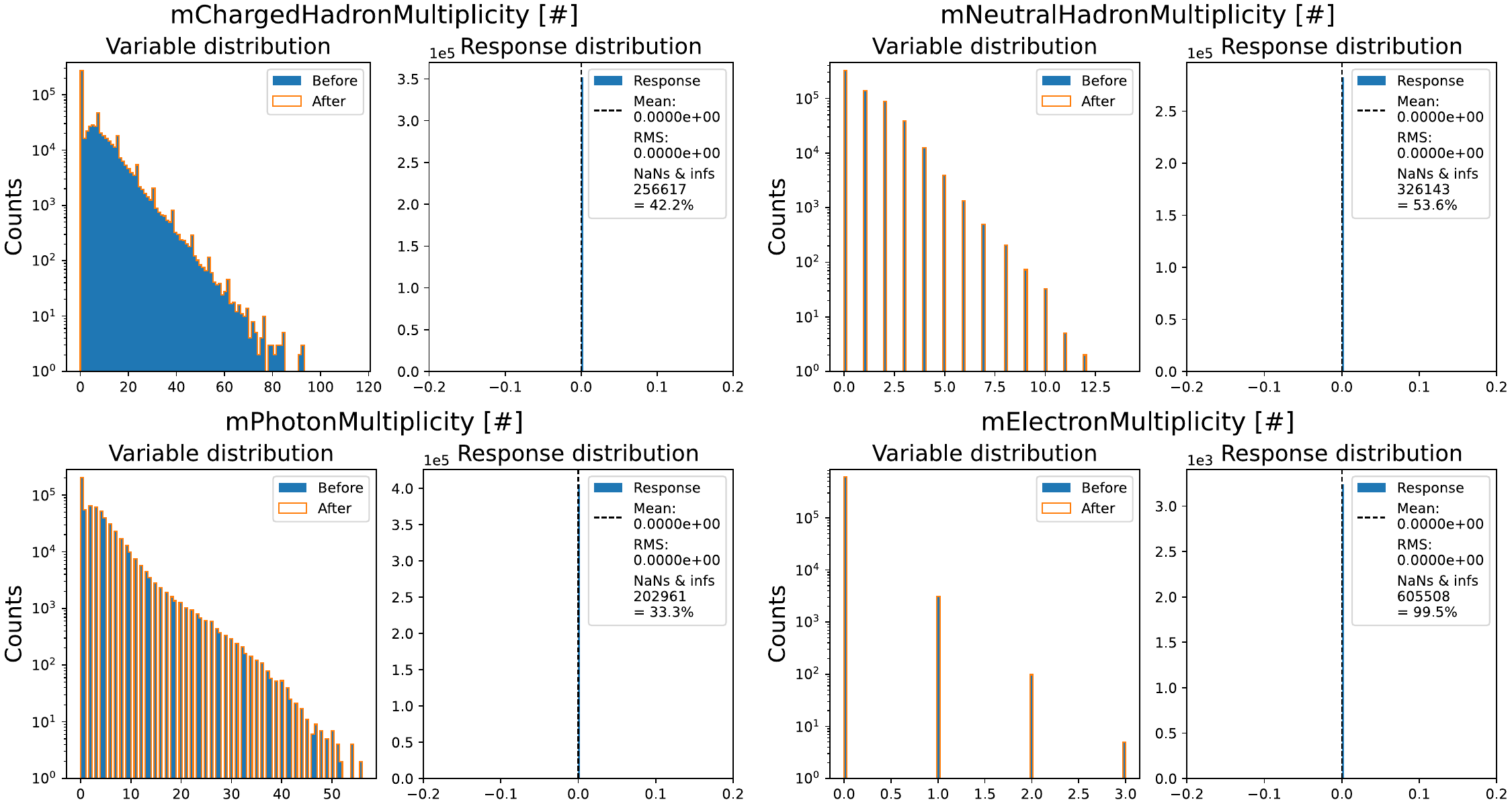}
    \includegraphics[page=5,width=0.85\linewidth]{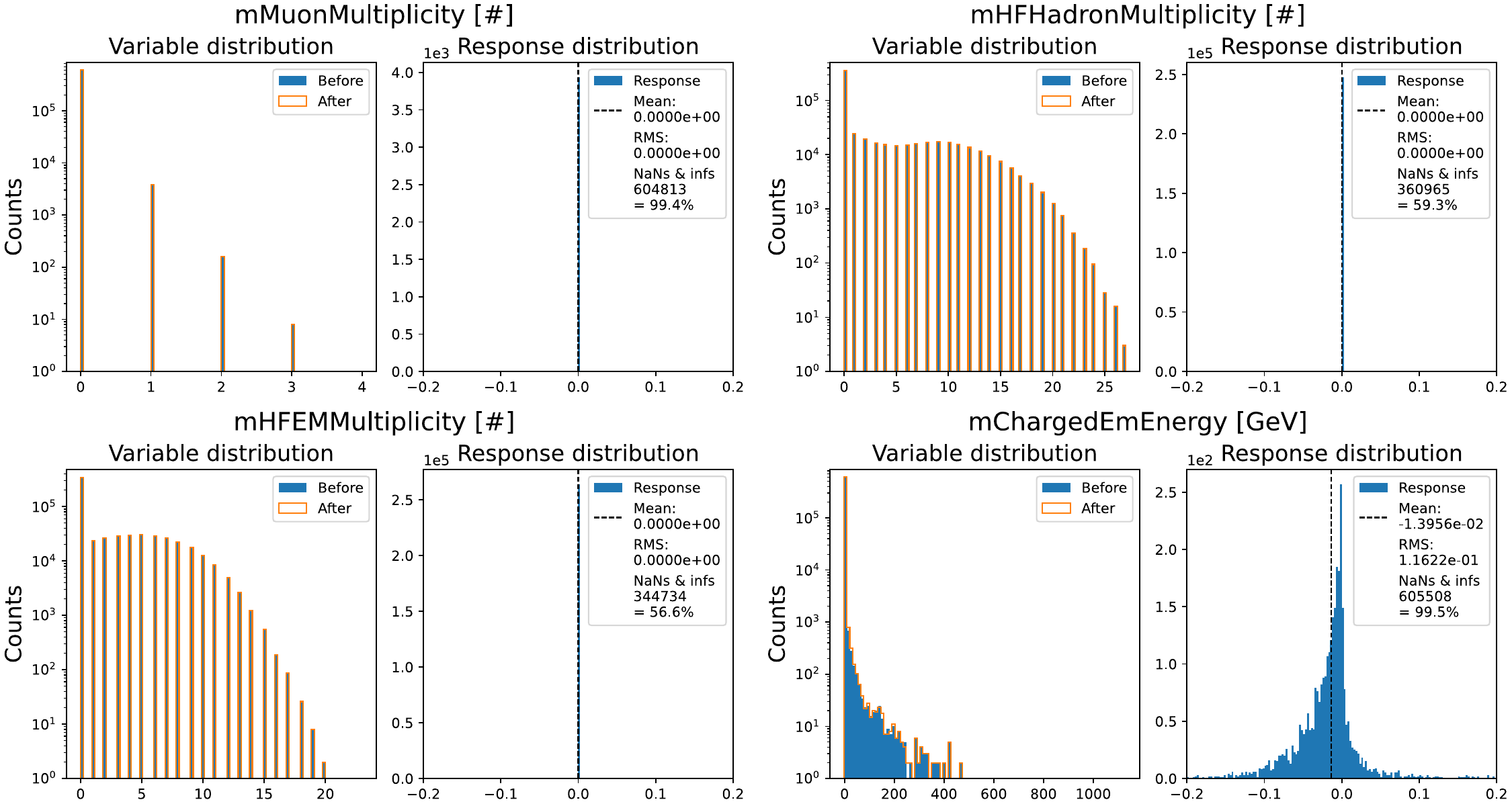}
    \includegraphics[page=6,width=0.85\linewidth]{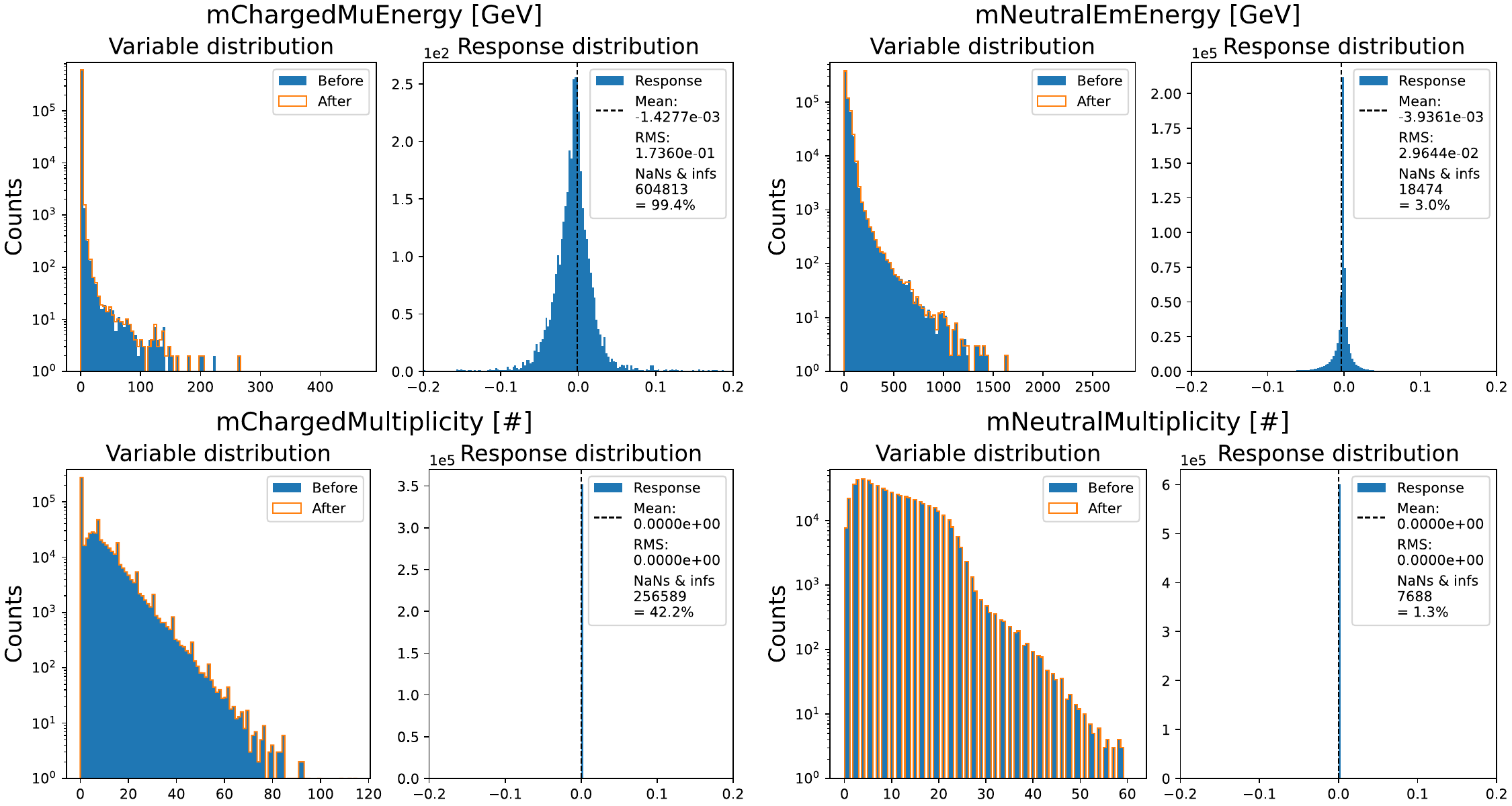}

    \caption{Original distribution and reconstructed distributions of twelve jet variables after compression with $R = 1.7$. Alongside each distribution is a histogram of the response for that variable.}
    \label{fig:beforeafter1.6x2}
\end{figure}

\begin{figure}[H]
    \centering
    \includegraphics[page=2,width=0.85\linewidth]{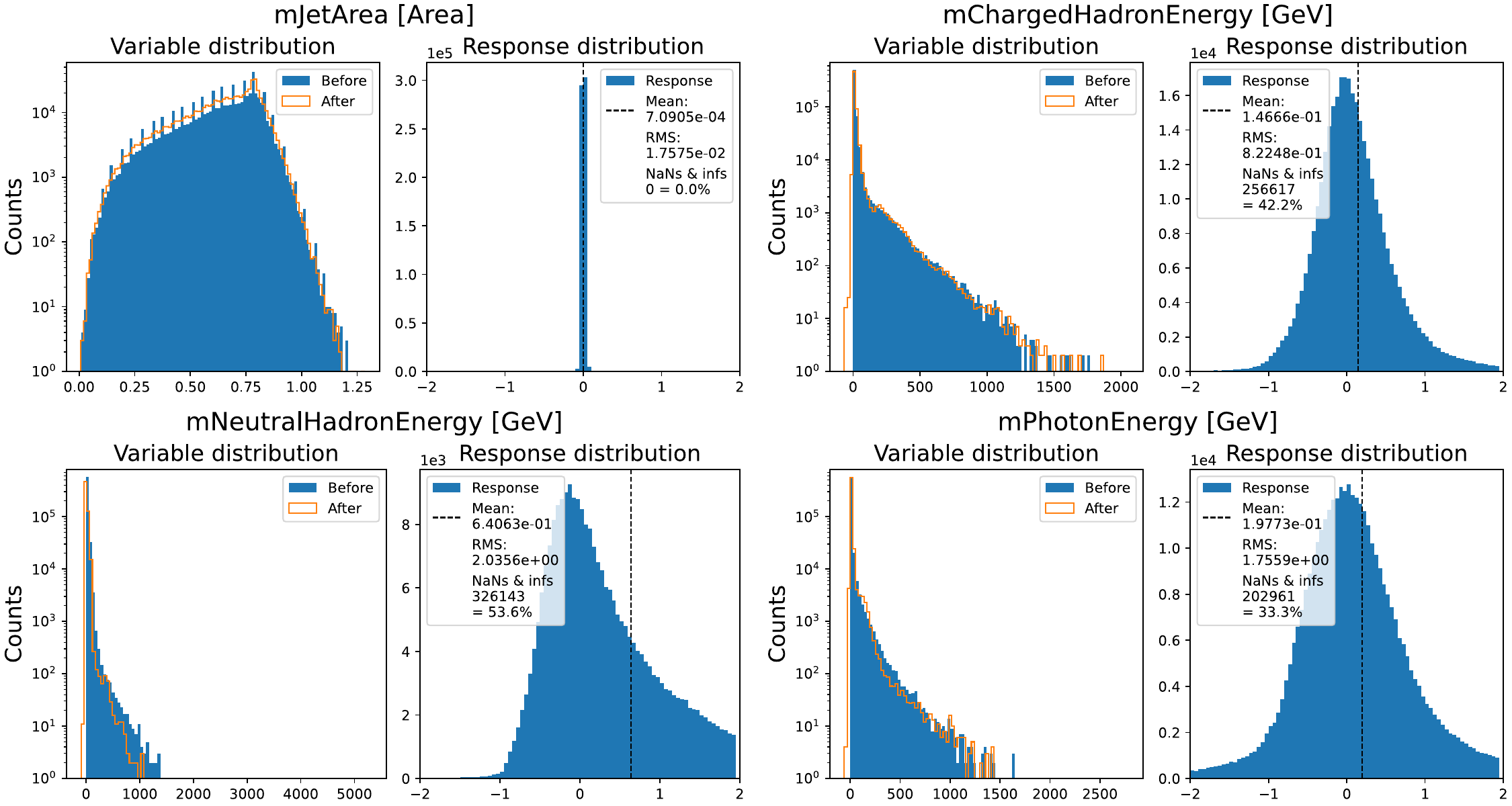}
    \includegraphics[page=3,width=0.85\linewidth]{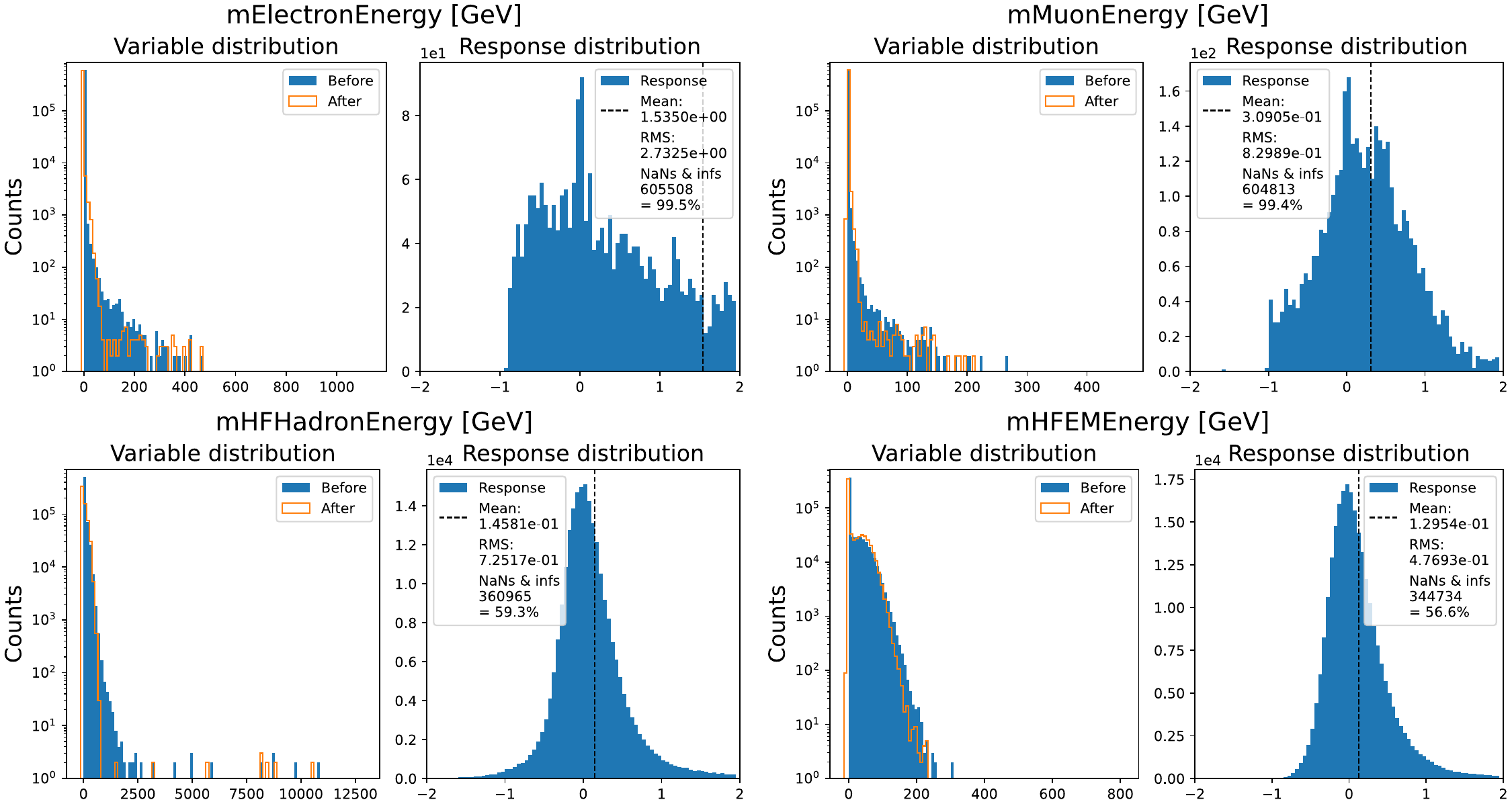}

    \caption{Original distribution and reconstructed distributions of eight jet variables after compression with $R=6$. Alongside each distribution is a histogram of the response for that variable.}
    \label{fig:beforeafter6x}
\end{figure}
\begin{figure}[H]
    \centering
    \includegraphics[page=4,width=0.85\linewidth]{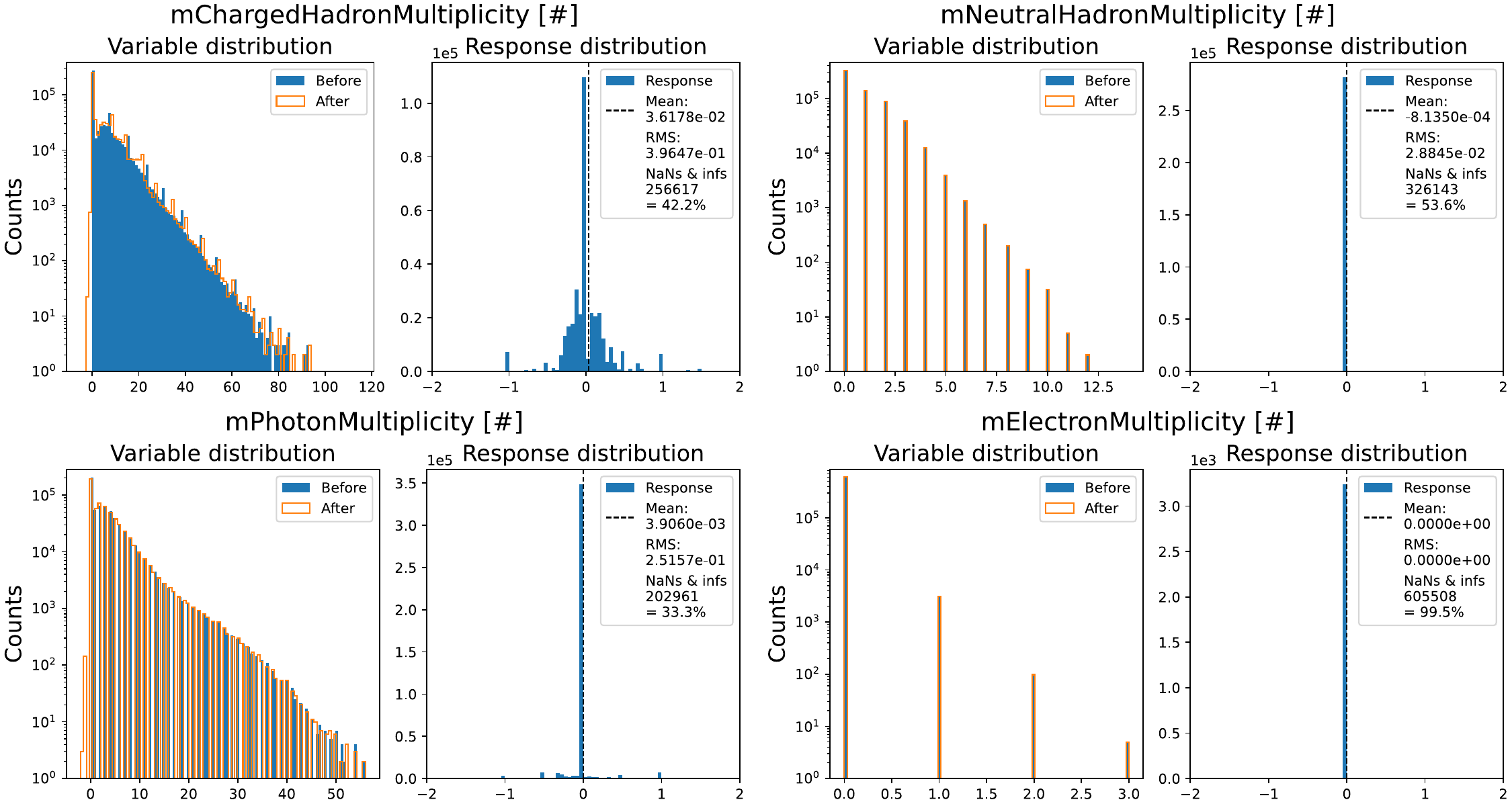}
    \includegraphics[page=5,width=0.85\linewidth]{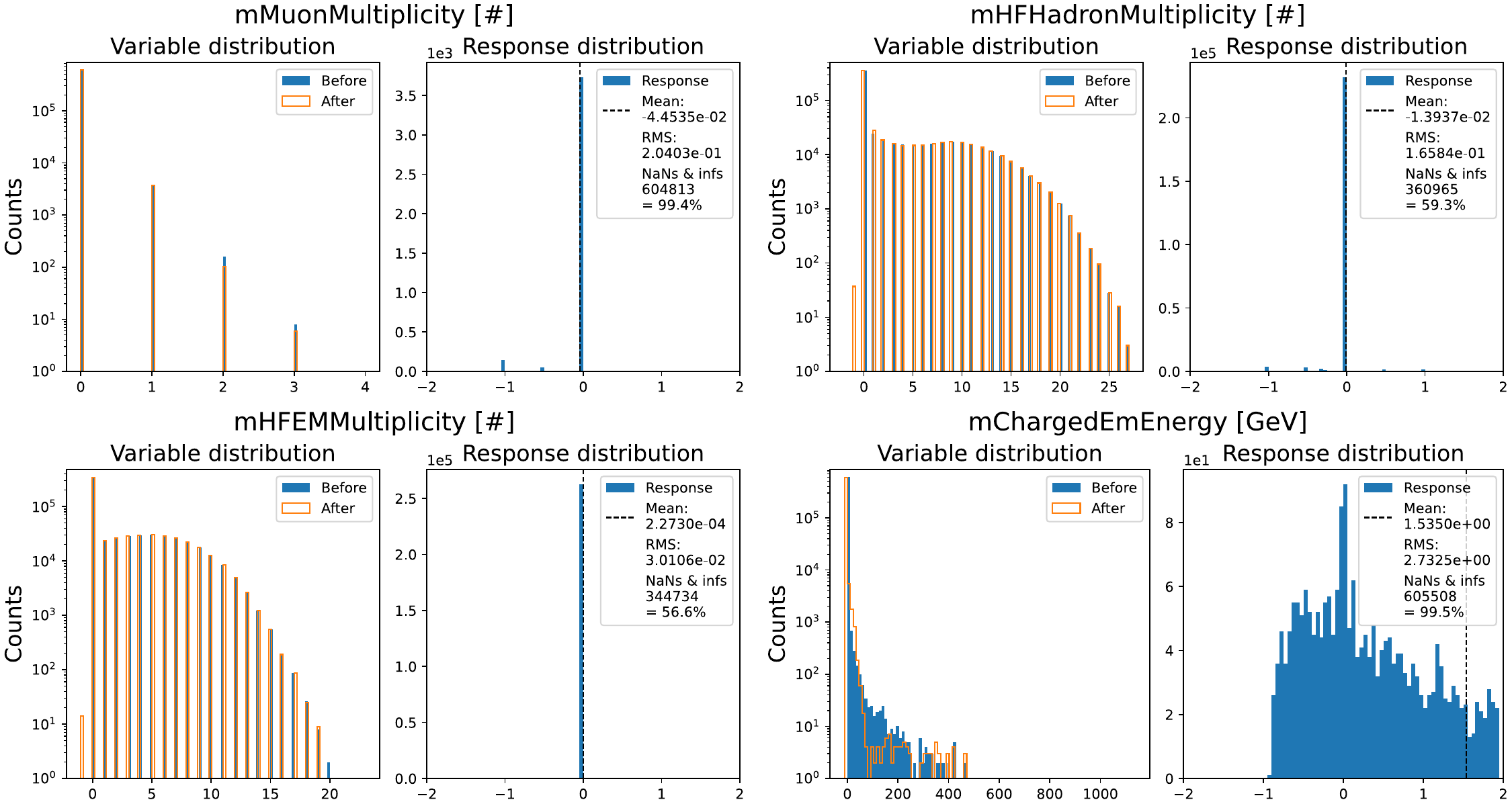}
    \includegraphics[page=6,width=0.85\linewidth]{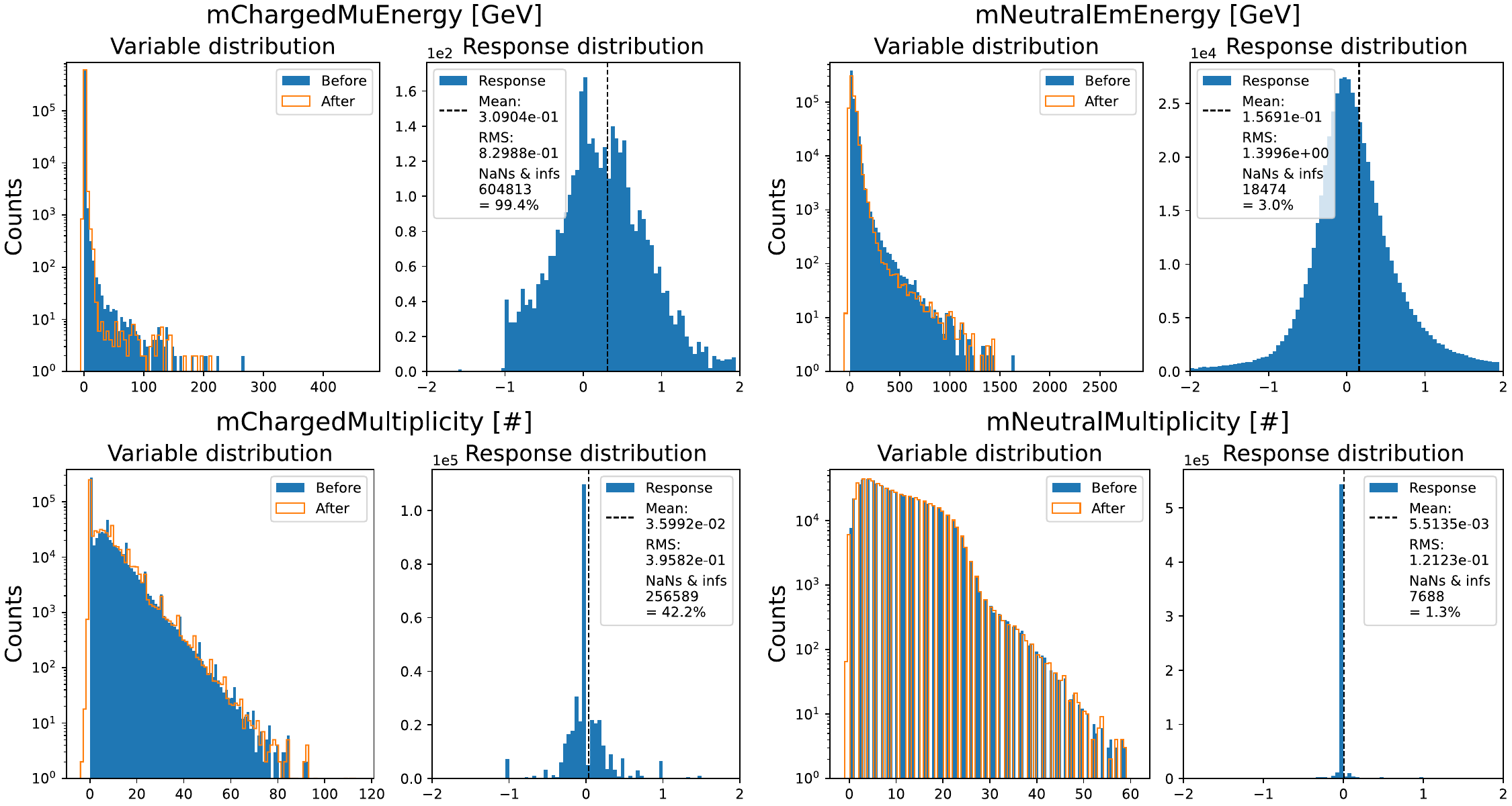}

    \caption{Original distribution and reconstructed distributions of twelve jet variables after compression with $R=6$. Alongside each distribution is a histogram of the response for that variable.}
    \label{fig:beforeafter6x2}
\end{figure}

\subsection{Variable evaluation metrics}
\label{lab:Results_table_appendix}
Here, the final averaged evaluation metrics of all variables are presented. Table \ref{tab:Appendix_results_table_1.6} shows the values for $R = 1.7$, while Table \ref{tab:Appendix_results_table_6x} shows it for $R = 6$.
\begin{table}[H]
    \centering
    \caption{Residual and Response distribution means and RMS values for all variables in the dataset. These values are presented at $R = 1.7$, and all values have been averaged over $5$ runs, with an added statistical error of two standard deviations.}
    \resizebox{\linewidth}{!}{\begin{tabular}{|c|c|c|c|c|}
        \hline
        \multirow{2}{*}{Variable ($R = 1.7$)} &  \multicolumn{2}{c|}{Response} & \multicolumn{2}{c|}{Residual}\\
        \cline{2-5}
            & Mean & RMS & Mean & RMS\\
        \hline
$p_T$& $\SI{-1.07e-03}{}\ \pm \SI{1.34e-02}{}$ & $\SI{2.09e-02}{} \pm \SI{3.56e-03}{}$ & $\SI{-1.44e-02}{} \pm \SI{1.04e-01}{}$ & $\SI{2.12e-01}{} \pm \SI{5.29e-02}{}$ \\ \hline 

$\eta$& $\SI{3.75e-04}{}\ \pm \SI{6.11e-04}{}$ & $\SI{8.12e-01}{} \pm \SI{1.17e+00}{}$ & $\SI{-1.12e-03}{} \pm \SI{2.67e-03}{}$ & $\SI{2.09e-03}{} \pm \SI{1.45e-03}{}$ \\ \hline 

$\phi$& $\SI{3.44e-04}{}\ \pm \SI{8.64e-04}{}$ & $\SI{1.93e-01}{} \pm \SI{4.32e-01}{}$ & $\SI{2.45e-04}{} \pm \SI{1.80e-03}{}$ & $\SI{9.91e-04}{} \pm \SI{1.12e-03}{}$ \\ \hline 

mass& $\SI{2.39e-01}{}\ \pm \SI{7.87e+00}{}$ & $\SI{4.38e+03}{} \pm \SI{4.47e+03}{}$ & $\SI{-8.05e-03}{} \pm \SI{2.51e-02}{}$ & $\SI{3.98e-02}{} \pm \SI{1.42e-02}{}$ \\ \hline 

mJetArea& $\SI{6.12e-05}{}\ \pm \SI{1.81e-04}{}$ & $\SI{3.13e-04}{} \pm \SI{1.48e-04}{}$ & $\SI{3.21e-05}{} \pm \SI{8.90e-05}{}$ & $\SI{1.10e-04}{} \pm \SI{5.77e-05}{}$ \\ \hline 

mChargedHadronEnergy& $\SI{1.58e-03}{}\ \pm \SI{1.70e-02}{}$ & $\SI{2.85e-02}{} \pm \SI{1.30e-02}{}$ & $\SI{1.68e-02}{} \pm \SI{1.43e-01}{}$ & $\SI{1.71e-01}{} \pm \SI{7.33e-02}{}$ \\ \hline 

mNeutralHadronEnergy& $\SI{7.05e-02}{}\ \pm \SI{9.88e-02}{}$ & $\SI{2.22e-01}{} \pm \SI{6.59e-02}{}$ & $\SI{2.77e-01}{} \pm \SI{5.23e-01}{}$ & $\SI{6.94e-01}{} \pm \SI{2.26e-01}{}$ \\ \hline 

mPhotonEnergy& $\SI{-2.75e-02}{}\ \pm \SI{7.48e-02}{}$ & $\SI{6.84e-02}{} \pm \SI{1.09e-01}{}$ & $\SI{-8.00e-02}{} \pm \SI{1.87e-01}{}$ & $\SI{1.52e-01}{} \pm \SI{1.77e-01}{}$ \\ \hline 

mElectronEnergy& $\SI{-7.71e-02}{}\ \pm \SI{1.05e-01}{}$ & $\SI{1.44e-01}{} \pm \SI{7.47e-02}{}$ & $\SI{1.71e-02}{} \pm \SI{5.32e-02}{}$ & $\SI{8.40e-02}{} \pm \SI{4.15e-02}{}$ \\ \hline 

mMuonEnergy& $\SI{1.29e-02}{}\ \pm \SI{1.97e-02}{}$ & $\SI{8.04e-02}{} \pm \SI{9.77e-02}{}$ & $\SI{1.18e-02}{} \pm \SI{1.46e-02}{}$ & $\SI{3.15e-02}{} \pm \SI{7.05e-03}{}$ \\ \hline 

mHFHadronEnergy& $\SI{-1.10e-02}{}\ \pm \SI{4.66e-02}{}$ & $\SI{1.77e-01}{} \pm \SI{2.48e-02}{}$ & $\SI{-3.15e-01}{} \pm \SI{1.07e+00}{}$ & $\SI{1.85e+00}{} \pm \SI{7.31e-01}{}$ \\ \hline 

mHFEMEnergy& $\SI{1.78e-03}{}\ \pm \SI{7.40e-03}{}$ & $\SI{1.41e-02}{} \pm \SI{3.63e-03}{}$ & $\SI{1.22e-02}{} \pm \SI{8.26e-02}{}$ & $\SI{6.93e-02}{} \pm \SI{5.54e-02}{}$ \\ \hline 

mChargedHadronMultiplicity& $\SI{-1.00e-03}{}\ \pm \SI{5.04e-03}{}$ & $\SI{4.48e-03}{} \pm \SI{4.90e-03}{}$ & $\SI{-3.13e-03}{} \pm \SI{1.82e-02}{}$ & $\SI{9.68e-03}{} \pm \SI{1.50e-02}{}$ \\ \hline 

mNeutralHadronMultiplicity& $\SI{-1.22e-04}{}\ \pm \SI{1.29e-03}{}$ & $\SI{8.76e-04}{} \pm \SI{9.42e-04}{}$ & $\SI{-1.19e-04}{} \pm \SI{1.51e-03}{}$ & $\SI{9.89e-04}{} \pm \SI{1.20e-03}{}$ \\ \hline 

mPhotonMultiplicity& $\SI{-1.14e-03}{}\ \pm \SI{3.62e-03}{}$ & $\SI{2.72e-03}{} \pm \SI{4.14e-03}{}$ & $\SI{-2.69e-03}{} \pm \SI{7.44e-03}{}$ & $\SI{4.92e-03}{} \pm \SI{7.12e-03}{}$ \\ \hline 

mElectronMultiplicity& $\SI{1.07e-03}{}\ \pm \SI{3.87e-03}{}$ & $\SI{2.37e-03}{} \pm \SI{2.37e-03}{}$ & $\SI{-1.54e-05}{} \pm \SI{9.96e-05}{}$ & $\SI{2.11e-04}{} \pm \SI{1.75e-04}{}$ \\ \hline 

mMuonMultiplicity& $\SI{1.12e-03}{}\ \pm \SI{1.22e-03}{}$ & $\SI{2.51e-03}{} \pm \SI{6.69e-04}{}$ & $\SI{5.67e-05}{} \pm \SI{1.16e-04}{}$ & $\SI{2.41e-04}{} \pm \SI{6.35e-05}{}$ \\ \hline 

mHFHadronMultiplicity& $\SI{-1.34e-03}{}\ \pm \SI{1.84e-03}{}$ & $\SI{2.53e-03}{} \pm \SI{1.94e-03}{}$ & $\SI{-2.67e-03}{} \pm \SI{3.33e-03}{}$ & $\SI{4.44e-03}{} \pm \SI{4.05e-03}{}$ \\ \hline 

mHFEMMultiplicity& $\SI{2.41e-04}{}\ \pm \SI{2.51e-03}{}$ & $\SI{1.98e-03}{} \pm \SI{1.33e-03}{}$ & $\SI{5.98e-04}{} \pm \SI{4.16e-03}{}$ & $\SI{3.08e-03}{} \pm \SI{2.95e-03}{}$ \\ \hline 

mChargedEmEnergy& $\SI{-7.72e-02}{}\ \pm \SI{1.05e-01}{}$ & $\SI{1.44e-01}{} \pm \SI{7.48e-02}{}$ & $\SI{1.72e-02}{} \pm \SI{5.30e-02}{}$ & $\SI{8.40e-02}{} \pm \SI{4.15e-02}{}$ \\ \hline 

mChargedMuEnergy& $\SI{1.29e-02}{}\ \pm \SI{1.97e-02}{}$ & $\SI{8.05e-02}{} \pm \SI{9.78e-02}{}$ & $\SI{1.18e-02}{} \pm \SI{1.46e-02}{}$ & $\SI{3.15e-02}{} \pm \SI{7.07e-03}{}$ \\ \hline 

mNeutralEmEnergy& $\SI{-1.73e-02}{}\ \pm \SI{5.42e-02}{}$ & $\SI{5.89e-02}{} \pm \SI{8.87e-02}{}$ & $\SI{-6.70e-02}{} \pm \SI{2.57e-01}{}$ & $\SI{1.75e-01}{} \pm \SI{1.81e-01}{}$ \\ \hline 

mChargedMultiplicity& $\SI{-9.83e-04}{}\ \pm \SI{5.04e-03}{}$ & $\SI{4.46e-03}{} \pm \SI{4.88e-03}{}$ & $\SI{-3.07e-03}{} \pm \SI{1.83e-02}{}$ & $\SI{9.74e-03}{} \pm \SI{1.51e-02}{}$ \\ \hline 

mNeutralMultiplicity& $\SI{-8.97e-04}{}\ \pm \SI{1.42e-03}{}$ & $\SI{1.56e-03}{} \pm \SI{1.93e-03}{}$ & $\SI{-5.36e-03}{} \pm \SI{7.37e-03}{}$ & $\SI{7.34e-03}{} \pm \SI{6.60e-03}{}$ \\ \hline 
    \end{tabular}}
    \label{tab:Appendix_results_table_1.6}
\end{table}

\begin{table}[H]
    \centering
    \caption{Residual and Response distribution means and RMS values for all variables in the dataset. These values are presented at $R = 6$, and all values have been averaged over $5$ runs, with an added statistical error of two standard deviations.}
    \resizebox{1\linewidth}{!}{\begin{tabular}{|c|c|c|c|c|}
        \hline
        \multirow{2}{*}{Variable ($R = 6$)} &  \multicolumn{2}{c|}{Response} & \multicolumn{2}{c|}{Residual}\\
        \cline{2-5}
            & Mean & RMS & Mean & RMS\\
        \hline
          $p_T$& $\SI{9.08e-02}{}\ \pm \SI{2.37e-02}{}$ & $\SI{3.67e-01}{} \pm \SI{4.17e-02}{}$ & $\SI{-5.60e-02}{} \pm \SI{1.52e-01}{}$ & $\SI{1.17e+01}{} \pm \SI{3.13e+00}{}$ \\ \hline 

$\eta$& $\SI{-5.42e-02}{}\ \pm \SI{3.34e-01}{}$ & $\SI{8.28e+01}{} \pm \SI{1.32e+02}{}$ & $\SI{-2.14e-03}{} \pm \SI{6.21e-03}{}$ & $\SI{1.47e-01}{} \pm \SI{3.67e-02}{}$ \\ \hline 

$\phi$& $\SI{1.14e-04}{}\ \pm \SI{1.52e-03}{}$ & $\SI{6.63e-01}{} \pm \SI{8.32e-01}{}$ & $\SI{2.52e-04}{} \pm \SI{1.46e-03}{}$ & $\SI{9.92e-03}{} \pm \SI{2.12e-02}{}$ \\ \hline 

mass& $\SI{-1.34e+01}{}\ \pm \SI{5.05e+01}{}$ & $\SI{5.95e+04}{} \pm \SI{3.42e+04}{}$ & $\SI{1.22e-02}{} \pm \SI{2.55e-02}{}$ & $\SI{1.86e+00}{} \pm \SI{1.94e-01}{}$ \\ \hline 

mJetArea& $\SI{2.77e-04}{}\ \pm \SI{9.83e-04}{}$ & $\SI{2.01e-02}{} \pm \SI{4.58e-03}{}$ & $\SI{-1.64e-05}{} \pm \SI{5.66e-04}{}$ & $\SI{1.16e-02}{} \pm \SI{2.13e-03}{}$ \\ \hline 

mChargedHadronEnergy& $\SI{1.29e-01}{}\ \pm \SI{2.32e-02}{}$ & $\SI{8.37e-01}{} \pm \SI{1.09e-01}{}$ & $\SI{-9.49e-02}{} \pm \SI{2.37e-01}{}$ & $\SI{1.58e+01}{} \pm \SI{2.09e+00}{}$ \\ \hline 

mNeutralHadronEnergy& $\SI{6.51e-01}{}\ \pm \SI{3.75e-02}{}$ & $\SI{2.02e+00}{} \pm \SI{1.35e-01}{}$ & $\SI{8.35e-02}{} \pm \SI{2.99e-01}{}$ & $\SI{1.74e+01}{} \pm \SI{1.09e+00}{}$ \\ \hline 

mPhotonEnergy& $\SI{2.02e-01}{}\ \pm \SI{1.38e-01}{}$ & $\SI{1.71e+00}{} \pm \SI{1.93e-01}{}$ & $\SI{-2.70e-01}{} \pm \SI{3.79e-01}{}$ & $\SI{1.52e+01}{} \pm \SI{1.51e+00}{}$ \\ \hline 

mElectronEnergy& $\SI{1.52e+00}{}\ \pm \SI{2.06e-01}{}$ & $\SI{2.70e+00}{} \pm \SI{2.67e-01}{}$ & $\SI{-3.08e-03}{} \pm \SI{2.08e-02}{}$ & $\SI{2.12e+00}{} \pm \SI{3.23e-01}{}$ \\ \hline 

mMuonEnergy& $\SI{3.53e-01}{}\ \pm \SI{1.13e-01}{}$ & $\SI{9.15e-01}{} \pm \SI{1.01e-01}{}$ & $\SI{-9.21e-03}{} \pm \SI{1.14e-02}{}$ & $\SI{7.81e-01}{} \pm \SI{1.85e-01}{}$ \\ \hline 

mHFHadronEnergy& $\SI{1.43e-01}{}\ \pm \SI{6.72e-02}{}$ & $\SI{7.15e-01}{} \pm \SI{1.64e-01}{}$ & $\SI{-1.99e-02}{} \pm \SI{3.71e-01}{}$ & $\SI{3.59e+01}{} \pm \SI{8.51e-01}{}$ \\ \hline 

mHFEMEnergy& $\SI{1.35e-01}{}\ \pm \SI{2.05e-02}{}$ & $\SI{4.91e-01}{} \pm \SI{4.87e-02}{}$ & $\SI{9.50e-02}{} \pm \SI{9.11e-02}{}$ & $\SI{8.70e+00}{} \pm \SI{7.63e-01}{}$ \\ \hline 

mChargedHadronMultiplicity& $\SI{2.60e-02}{}\ \pm \SI{2.32e-02}{}$ & $\SI{3.59e-01}{} \pm \SI{7.28e-02}{}$ & $\SI{7.48e-03}{} \pm \SI{3.39e-02}{}$ & $\SI{1.14e+00}{} \pm \SI{3.21e-01}{}$ \\ \hline 

mNeutralHadronMultiplicity& $\SI{6.98e-03}{}\ \pm \SI{2.29e-02}{}$ & $\SI{7.91e-02}{} \pm \SI{9.70e-02}{}$ & $\SI{7.03e-04}{} \pm \SI{2.75e-03}{}$ & $\SI{1.09e-01}{} \pm \SI{9.42e-02}{}$ \\ \hline 

mPhotonMultiplicity& $\SI{4.37e-03}{}\ \pm \SI{1.44e-02}{}$ & $\SI{2.24e-01}{} \pm \SI{6.04e-02}{}$ & $\SI{-5.24e-03}{} \pm \SI{2.14e-02}{}$ & $\SI{4.71e-01}{} \pm \SI{1.12e-01}{}$ \\ \hline 

mElectronMultiplicity& $\SI{8.09e-04}{}\ \pm \SI{3.07e-03}{}$ & $\SI{2.06e-02}{} \pm \SI{1.90e-02}{}$ & $\SI{-4.56e-05}{} \pm \SI{1.33e-04}{}$ & $\SI{1.76e-03}{} \pm \SI{1.56e-03}{}$ \\ \hline 

mMuonMultiplicity& $\SI{-2.05e-02}{}\ \pm \SI{4.75e-02}{}$ & $\SI{1.10e-01}{} \pm \SI{1.77e-01}{}$ & $\SI{1.68e-05}{} \pm \SI{1.77e-04}{}$ & $\SI{1.10e-02}{} \pm \SI{1.81e-02}{}$ \\ \hline 

mHFHadronMultiplicity& $\SI{-1.93e-02}{}\ \pm \SI{2.89e-02}{}$ & $\SI{1.56e-01}{} \pm \SI{1.57e-02}{}$ & $\SI{6.87e-04}{} \pm \SI{4.38e-03}{}$ & $\SI{2.16e-01}{} \pm \SI{5.30e-02}{}$ \\ \hline 

mHFEMMultiplicity& $\SI{6.34e-04}{}\ \pm \SI{3.67e-03}{}$ & $\SI{5.84e-02}{} \pm \SI{2.48e-02}{}$ & $\SI{4.87e-04}{} \pm \SI{7.10e-03}{}$ & $\SI{9.81e-02}{} \pm \SI{2.35e-02}{}$ \\ \hline 

mChargedEmEnergy& $\SI{1.52e+00}{}\ \pm \SI{2.06e-01}{}$ & $\SI{2.70e+00}{} \pm \SI{2.67e-01}{}$ & $\SI{-3.17e-03}{} \pm \SI{2.07e-02}{}$ & $\SI{2.12e+00}{} \pm \SI{3.23e-01}{}$ \\ \hline 

mChargedMuEnergy& $\SI{3.53e-01}{}\ \pm \SI{1.13e-01}{}$ & $\SI{9.15e-01}{} \pm \SI{1.01e-01}{}$ & $\SI{-9.16e-03}{} \pm \SI{1.13e-02}{}$ & $\SI{7.81e-01}{} \pm \SI{1.85e-01}{}$ \\ \hline 

mNeutralEmEnergy& $\SI{1.64e-01}{}\ \pm \SI{8.33e-02}{}$ & $\SI{1.38e+00}{} \pm \SI{1.51e-01}{}$ & $\SI{-8.00e-02}{} \pm \SI{3.08e-01}{}$ & $\SI{1.76e+01}{} \pm \SI{1.44e+00}{}$ \\ \hline 

mChargedMultiplicity& $\SI{2.59e-02}{}\ \pm \SI{2.32e-02}{}$ & $\SI{3.58e-01}{} \pm \SI{7.34e-02}{}$ & $\SI{7.50e-03}{} \pm \SI{3.39e-02}{}$ & $\SI{1.14e+00}{} \pm \SI{3.21e-01}{}$ \\ \hline 

mNeutralMultiplicity& $\SI{4.23e-03}{}\ \pm \SI{6.92e-03}{}$ & $\SI{9.32e-02}{} \pm \SI{4.73e-02}{}$ & $\SI{-2.91e-03}{} \pm \SI{2.04e-02}{}$ & $\SI{4.32e-01}{} \pm \SI{1.17e-01}{}$ \\ \hline 
    \end{tabular}}
    \label{tab:Appendix_results_table_6x}
\end{table}

\subsection{Comparison to Zip}
\label{app:baler_vs_zip}

\begin{figure}
    \centering
    \includegraphics[width=1\linewidth]{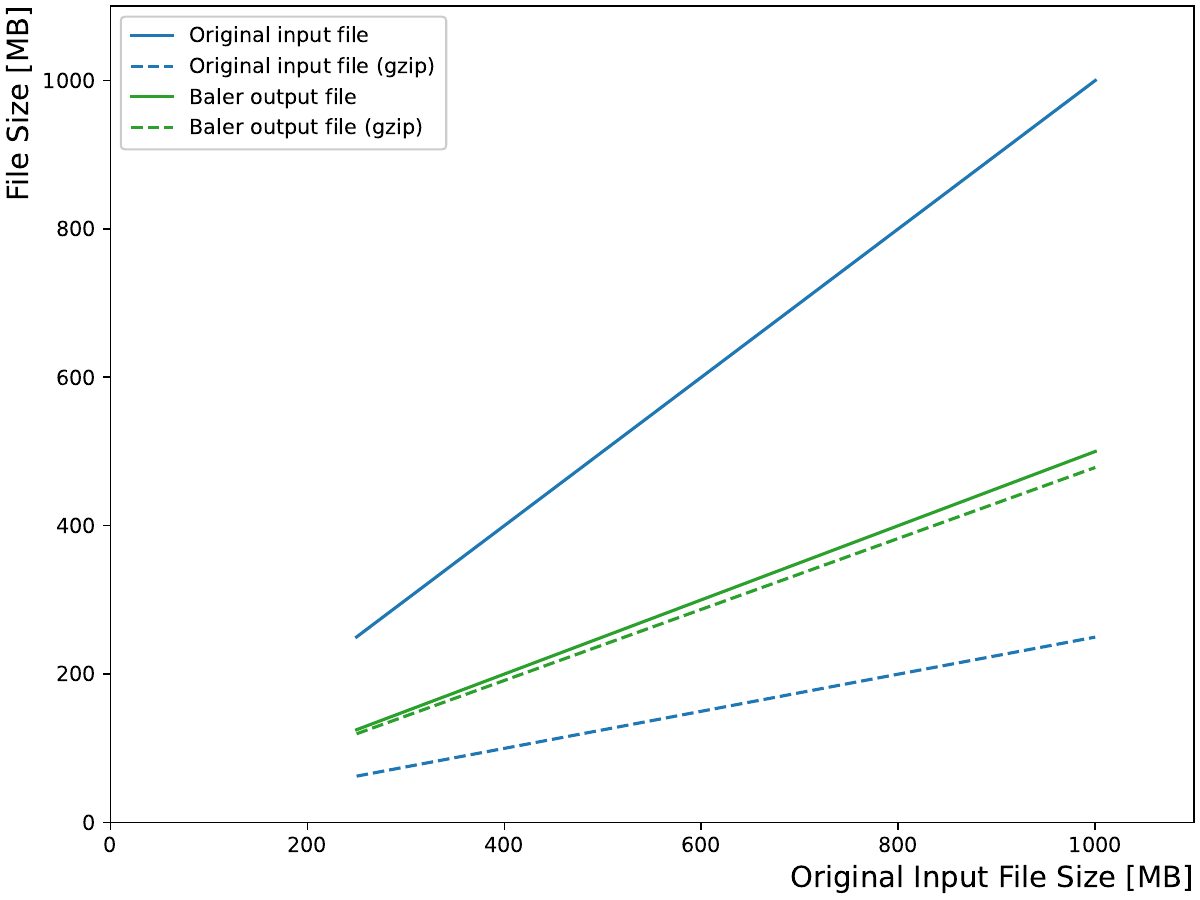}
    \caption{The output file size of the different scenarios tested as a function of the input file size. Note that in this scenario, Baler used a $R = 2.0$}
    \label{fig:zip}
\end{figure}

To make sure that relationship between the zipped Baler output and zipped input is not dependent on file size, we tested this for different input file sizes: 250, 500, 750, 1000, 1250, and 1500 MB. All originating from the same data source~\cite{dataset}. Figure \ref{fig:zip} shows the output file size of the different scenarios tested as a function of the input file size.

\end{document}